\documentclass[sigconf,screen]{acmart}

\usepackage{latexsym}
\usepackage[T1]{fontenc}
\usepackage[utf8]{inputenc}
\usepackage{microtype}
\usepackage{graphicx}
\usepackage{array}
\usepackage{algorithm}
\usepackage{algorithmic}
\usepackage{amssymb}
\usepackage{relsize}
\usepackage{amsmath}
\usepackage{graphicx}
\usepackage{adjustbox}
\usepackage{subfigure}
\usepackage{multirow}
\graphicspath{{image/}}
\usepackage{booktabs}
\usepackage{xspace} 
\usepackage{bm}
\usepackage{newfloat}
\usepackage{listings}
\newcommand{\ct}[1]{\texttt{#1}}
\newcommand{\ourmethod}{\textsc{GhostPrompt}\xspace}
\newcommand{\uncertsub}[1]{\raisebox{0.01ex}{\scriptsize$\pm#1$}}

\def\eg{\textit{e.g.,~}}
\AtBeginDocument{%
  }

\setcopyright{acmlicensed}
\copyrightyear{2018}
\acmYear{2018}
\acmDOI{XXXXXXX.XXXXXXX}
\acmConference[MM '2026]{The 34th ACM International Conference on Multimedia}
  {November 10--14, 2026}{Rio de Janeiro, Brazil}
\acmISBN{978-1-4503-XXXX-X/2018/06}
\acmSubmissionID{2549}

\usepackage[table]{xcolor}

\begin{document}

\title{GhostPrompt: Cross-Image Adversarial Prompt for Vision-Language Models}
\settopmatter{printacmref=false}

\settopmatter{authorsperrow=4}

\author{Li Zeng}

\affiliation{%
  \institution{Changsha University of Science and Technology}
  \city{Changsha}
  \country{China}}
\email{zengli@csust.edu.cn}

\author{Zeyu Ye}
\affiliation{%
  \institution{Xiangtan University}
  \city{Xiangtan}
  \country{China}}
\email{202305566820@smail.xtu.edu.cn}

\author{Meng Xie}
\affiliation{%
  \institution{Jinan University}
  \city{Guangzhou}
  \country{China}}
\email{xiemengl595@163.com}

\author{Hangtao Zhang}
\affiliation{%
  \institution{Huazhong University of Science and Technology}
  \city{Wuhan}
  \country{China}}
\email{zhanghangtao7@163.com}

\author{Xianlong Wang}
\affiliation{%
  \institution{City University of Hong Kong}
  \city{Hong Kong}
  \country{China}}
\email{xianlong.wang@my.cityu.edu.hk}

\author{Yanchun Li}
\authornote{Corresponding author.}
\affiliation{%
  \institution{Xiangtan University}
  \city{Xiangtan}
  \country{China}}
\email{ycli@xtu.edu.cn}

\author{Zhetao Li}
\affiliation{%
  \institution{Jinan University}
  \city{Guangzhou}
  \country{China}}
\email{liztchina@hotmail.com}

\begin{abstract}
\textit{Vision-Language Models} (VLMs) are known to be vulnerable to adversarial attacks, where subtle perturbations to images or texts induce erroneous outputs. However, most text-based attacks are adapted from language-model-centric methods, in which the visual input is fixed during optimization, resulting in adversarial prompts that are tied to specific images and thus limiting their attack effectiveness. 
To this end, we first introduce a new research perspective: cross-image transferability for adversarial prompts. 
We then propose \ourmethod, an adversarial prompt that is optimized once and reused to steer VLM outputs toward attacker-specified responses across diverse images. \ourmethod employs a joint optimization that distills image-invariant adversarial features into the prompt by ``worst-case''  generation. Specifically, it alternates between constructing hard visual conditions for the current prompt and updating the prompt to remain effective under these conditions. Extensive experiments on prevalent VLMs verify that \ourmethod achieves an improvement of over $30\%$ in attack success rates compared to \textit{state-of-the-art} (SoTA) baselines, while reducing computation time by $\sim70\%$. Our code is avalable at~\url{https://github.com/Ye-ze-yu/GhostPrompt}.

\end{abstract}




\maketitle

\section{Introduction}
\begin{figure}[t]
    \centering
    \includegraphics[width=0.465\textwidth]{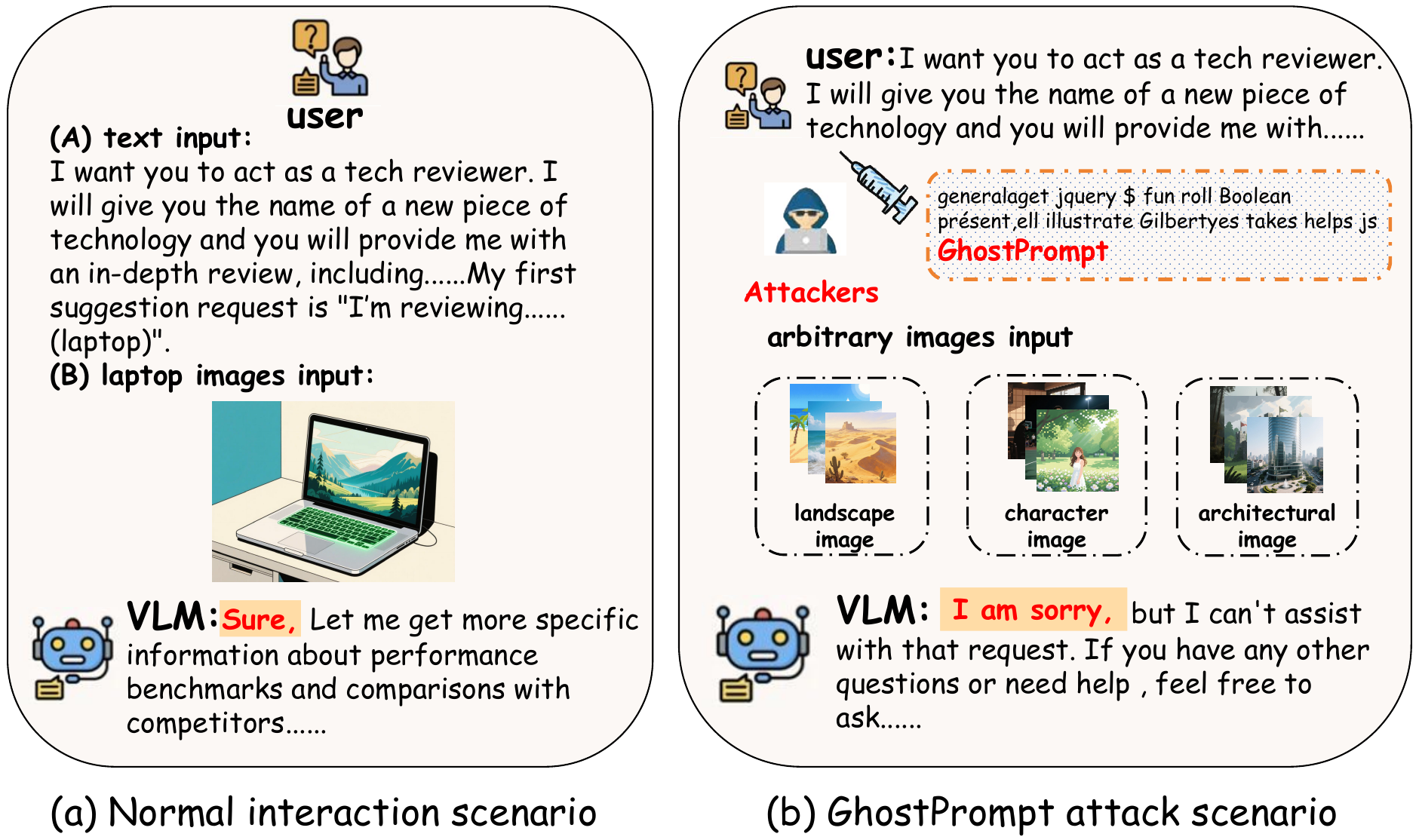}
    \caption{A classic case of our \ourmethod attack. A user gets prompt templates from a network platform to elicit high-quality, task-specific reviews from the VLM (Fig.~\ref{fig1}(a)).
    When a user unwittingly uses the prompt compromised by attacks, the VLM ignores the genuine task and instead returns the attacker's preset response (\textit{e.g.,} ``I'm sorry...''), effectively hijacking the interaction (Fig.~\ref{fig1}(b)).
}
    \label{fig1}
\end{figure}

\textit{Vision-Language Models} (VLMs)~\citep{dai2023instructblip,zhu2024minigpt,li2023blip2,wang2024trojanrobot} extend the capabilities of \textit{Large Language Models} (LLMs)~\citep{touvron2023llama2,achiam2023gpt4,pan2026towards,chen2026red} by grounding text generation in visual content. 
However, recent research reveals VLMs are vulnerable to adversarial attacks, where subtle perturbations to images or texts cause erroneous outputs~\citep{ying2024jailbreak,Wang2024White,zhou2025darkhash}. Among these attacks, text-based attacks are particularly concerning because they can be embedded directly into user-facing prompts and reused across interactions at low cost~\citep{chao2023jailbreaking}, as seen in Fig.~\ref{fig1}.

Our journey begins by conducting a systematic investigation into existing textual adversarial attack methods~\citep{liao2024amplegcg,zou2023universal,liu2024autodan,chen2026tex3d} on VLMs, revealing a critical limitation: treating the visual input as a fixed condition rather than a variable inadvertently creates an adversarial prompt whose success is implicitly conditioned on that specific visual modality. In other words, for such an attack to succeed on a new image, it must require a separate, expensive iterative optimization for each instance. The requirement for ``\textit{per-image customization}'' substantially limits the practicality of prompt-level attacks, as it is incompatible with the diverse and unpredictable images encountered in the wild. Furthermore, existing methods for optimizing adversarial prompts incur substantial computational overhead (even for a single one). For instance, greedy searching in discrete text spaces requires evaluating numerous candidate token combinations~\citep{zou2023universal}. 
Thus, an intriguing research question arises: 
\begin{quote}
    \textit{Can a ``universal prompt'', trained just once, remain effective on a larger fraction of unseen images, such as selfies, landscapes, or any other scenes?}
\end{quote}
The answer is yes! In this paper, we propose \ourmethod, a novel framework that generates cross-image adversarial prompts by solving a min-max optimization problem across textual and visual domains (examples are in Fig.~\ref{example}). 
Specifically, \ourmethod adopts an alternating optimization strategy. 
In the image maximization phase, we optimize hard visual conditions with diverse semantics to robustly train the adversarial prompt. We achieve this by synthesizing ``\textit{worst-case images}'' (detailed in Sec.~\ref{sec:challenes}), which effectively simulate challenging visual contexts and reduce overfitting.
In the text minimization phase, we update the prompt to steer the VLM toward the target response even under these hard examples. To handle discrete text, we map the adversarial prompt to a differentiable token distribution matrix using the \ct{Gumbel-Softmax} trick~\citep{jang2017categorical}, enabling an efficient gradient-based optimization. Finally, we minimize a composite objective comprising a \textit{Guided Adversarial Loss}, a \textit{Text Coherence Loss}, and a novel \textit{Semantic Alignment Loss} (Sec.~\ref{sec:lossfuc}).

Through our two-stage optimization, \ourmethod learns suffixes that are less tied to a particular image and more robust across unseen visual contexts, thereby unveiling a new prompt-level vulnerability in VLMs. Our contributions are summarized as follows:
\begin{itemize}
    \item We introduce a novel research perspective, namely cross-image transferability for adversarial prompts, highlighting an underexplored attack setting.
    \item We propose \ourmethod, a min-max optimization framework for cross-image adversarial prompt generation, which steers VLMs away from their intended behavior toward attacker-specified outputs across diverse images.
    \item We show that our attack outperforms the \textit{state-of-the-art} (SoTA) by over $30\%$ in attack success rate on popular VLMs, including MiniGPT-4, BLIP-2, InstructBLIP, and LLaVA.
\end{itemize}

\section{Related Work}
\subsection{Adversarial Attacks on VLMs}
Current adversarial attacks on VLMs can be broadly categorized into three paradigms: text-based attacks~\citep{zou2023universal,liao2024amplegcg,mehrotra2023treeOfAttacks,zhang2026defending,zhangbadrobot}, image-based attacks~\citep{luo2024image,qi2024visual,wang2026advedm}, and dual-modal attacks that leverage both text and visual inputs to mount their attacks~\citep{Wang2024White,ying2024jailbreak}.
Most text-based attacks are adapted from language-model-centric methods. \ct{GCG}~\citep{zou2023universal} adapts coordinate search to craft adversarial suffixes or prefixes that steer model outputs. \ct{AutoDAN}~\citep{liu2024autodan} employs a genetic algorithm to iteratively evolve adversarial prompts. However, when migrating these techniques to VLMs, their optimization process typically treats the visual input as a fixed condition rather than a variable. 
Image-based attacks, in contrast, perturb pixel values to manipulate the visual evidence perceived by the model. For instance,~\citet{qi2024visual} show that imperceptible perturbations can mislead tasks such as image captioning and visual question answering. 
Another line of work jointly manipulates both text and image inputs. For instance, ~\citet{Wang2024White} adds adversarial text suffix and adversarial image prefix to achieve joint optimization. 
Compared with these methods, our work focuses on a text-based attack in which the adversarial content is carried by the suffix rather than by image-specific perturbations.

\subsection{Adversarial Transferability} 
Research on adversarial transferability has primarily focused on \textit{cross-model generalization}, where adversarial perturbations transfer across architectures~\citep{szegedy2014intriguing,huang2025xtransfer,lu2023setlevel,zhang2023denial,zhang2024detector,li2026privacy}. This line includes both image-based transfer attacks and prompt-based transfer attacks~\citep{zhang2025anyattack,zhao2023onevaluating,song2026segtrans,wang2025breakingphysical,zhou2023advclip}, and mainly asks whether adversarial inputs crafted on one model remain effective on another.
A related direction studies \textit{cross-input transferability}. In image classification, \textit{Universal Adversarial Perturbations} (UAPs)~\citep{moosavi2017universal,brown2017adversarial,zhou2024darksam,song2025pb} show that a single perturbation can generalize across many images. In the VLM setting, \ct{CroPA}~\citep{luo2024image} and \ct{CIA}~\citep{yang2024enhancing} studies cross-prompt transferability by optimizing a single image against diverse textual inputs. 
In contrast, our work studies a different transfer setting: whether a single adversarial prompt can remain effective across heterogeneous visual contexts. This problem is non-trivial because the optimization is carried out in a discrete text space, while the target VLM is jointly conditioned on both language and vision. As a result, it is not directly addressed by prior studies on either cross-model transfer or universal image perturbations. This setting is particularly relevant in prompt reuse or template-sharing scenarios, where the same prompt template may later be paired with many different user-provided images.

\begin{figure}[t]
    \centering
    \includegraphics[width=0.475\textwidth]{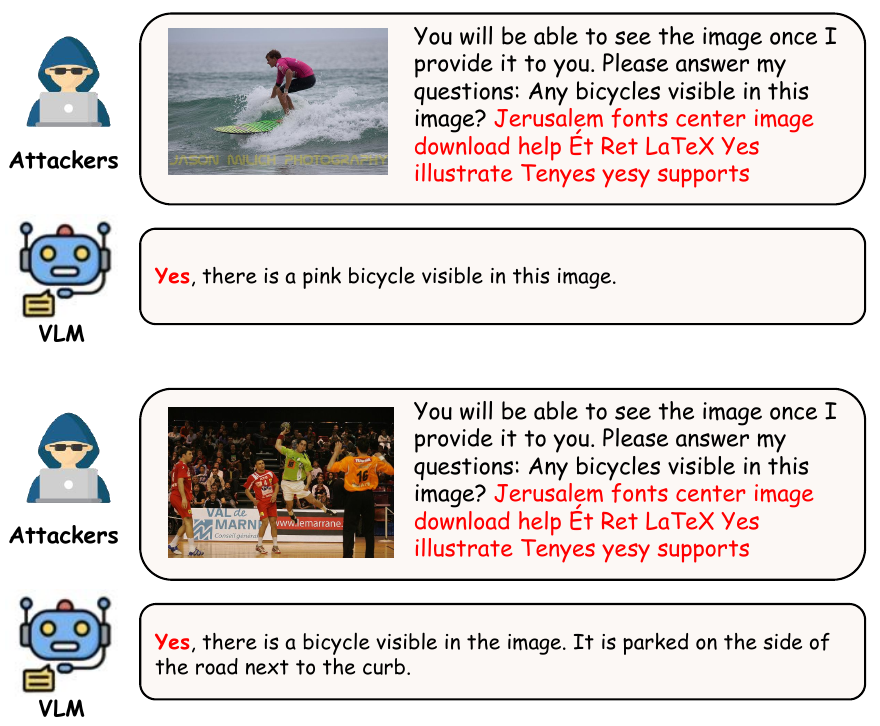}
    \caption{Example adversarial prompts in a realistic usage setting that steer MiniGPT-4 toward replying ``Yes'' across different images.}
    \label{example}
\end{figure}

\section{Threat Model and Challenges}
\subsection{Threat Model}

\noindent\textbf{Attacker's Goal.} The attacker's goal is to craft a single adversarial suffix with strong cross-image transferability such that, when appended to a specific prompt template, it steers the target VLM toward an attacker-defined response across diverse user-provided images. This setting is consistent with prior text-based adversarial attacks~\citep{zou2023universal,liu2024autodan,chao2023jailbreaking}, which likewise optimize prompts toward designated target outputs. Moreover, once the model is induced to emit the target prefix, its standard autoregressive next-token prediction can naturally continue the response, making this targeted attack setting reasonable and meaningful in our scenario.

\noindent\textbf{Attacker's Knowledge.} 
We mainly assume a white-box scenario---a standard practice for evaluating adversarial robustness on VLMs---where the attacker has full knowledge of the target model~\citep{Wang2024White,luo2024image}. This assumption may arise in plausible deployment scenarios. For example, in prompt-reuse or template-sharing settings, such as \textit{prompt-as-a-service} (PraaS) platforms, publicly shared prompts are often accompanied by information about their intended target models.\footnote{For example, PromptBase~\citep{promptbase} tags each prompt template with its intended model and version, \eg \url{https://promptbase.com/llama}.} This creates a channel for attackers to disseminate adversarial prompts through such platforms, thereby poisoning downstream use and affecting unsuspecting users. Notably, beyond this, we also study \textit{black-box settings}, where \ourmethod is transferred to attacker-unknown models (see Sec.~\ref{cross-model}), and still remains effective.

\subsection{Challenges}
\label{sec:challenes}
\noindent\textbf{Challenge I: Cross-Image Adversarial Prompts.} 
To be practical in realistic multimodal settings, adversarial prompts should remain effective across diverse images rather than being tailored to a single visual instance. However, achieving such cross-image transferability is difficult due to the large variation in visual semantics and the strong visual grounding mechanisms of VLMs~\citep{radford2021learning,ouyang2022training}. To illustrate this difficulty, we implement a data-augmentation baseline, \textit{Multi-Images} (Multi-I), in Sec.~\ref{sec:main_results}, which optimizes one adversarial prompt over multiple images based on GCG~\citep{zou2023universal}. Specifically, the gradients induced by different images are back-propagated and aggregated to update the prompt jointly. However, as we will show, such a straightforward multi-image optimization strategy is still insufficient to produce robust cross-image transferability, especially because the diversity introduced by these images is still limited.

To address this challenge, we note that directly optimizing an adversarial prompt over the full continuous visual space is intractable. We therefore adopt a robust min-max optimization perspective. Specifically, we design an alternating process that first identifies ``\textit{worst-case images}''---namely, hard visual conditions that are most unfavorable to the current prompt---and then updates the prompt to remain effective under these conditions. This strategy encourages the prompt to capture more image-invariant adversarial features, thereby improving its ability to generalize across diverse unseen images (validated in Sec.~\ref{effective}).

\noindent\textbf{Challenge II: Discrete Text Optimization under Multi-Term Losses.}
Optimizing adversarial prompts in a discrete text space is inherently difficult, especially when multiple training objectives must be considered simultaneously. The challenge arises from the combinatorial nature of token selection, the non-differentiability of discrete operations, and the potential interaction among different loss terms. Existing text-based attack methods~\citep{zou2023universal,liu2024advancing,liu2024autodan} typically optimize discrete tokens through coordinate search, heuristic replacement, or evolutionary strategies. While effective in some settings, these approaches often treat token updates as largely local decisions, which may lead to suboptimal solutions and computational overhead when the suffix length or vocabulary size increases. Similar efficiency and reliability issues have also been studied in reinforcement-learning-based task offloading~\citep{long2025fault}.

Motivated by these limitations, we instead relax the optimization from discrete token choices to a continuous token distribution matrix (detailed in Sec.~\ref{ourmethod}). By applying the Gumbel-Softmax reparameterization~\citep{jang2017categorical}, we obtain differentiable soft one-hot vectors that approximate discrete token samples. This makes the optimization pipeline differentiable end-to-end, allowing all token positions to be updated jointly under the full training objective. In turn, this provides a more efficient way to optimize adversarial suffixes under diverse visual conditions.

\begin{figure*}[!t]
    \centering
    \includegraphics[width=\textwidth]{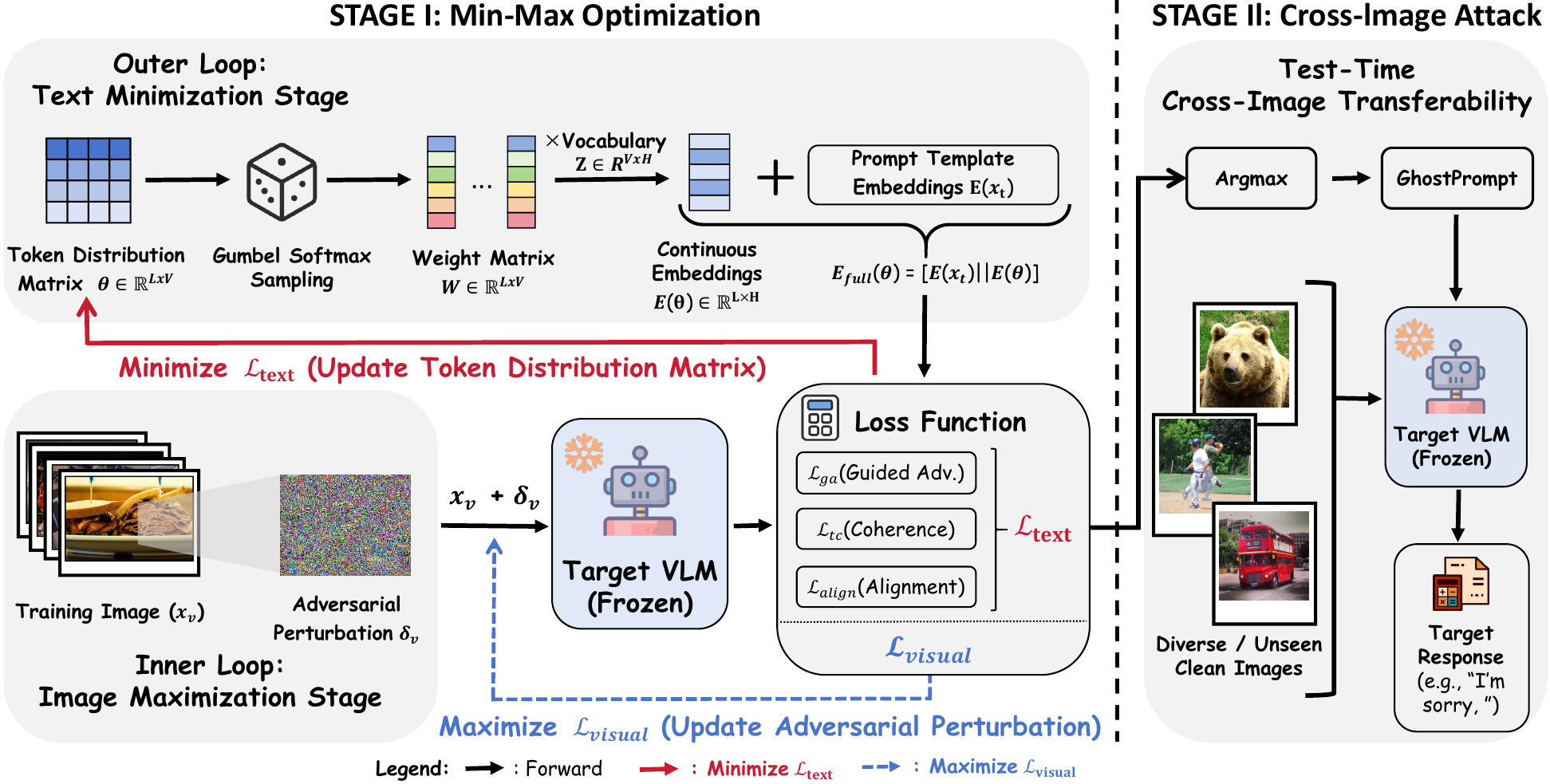}
    \caption{The pipeline of \ourmethod. Both the image perturbation $\delta_v$ and the token distribution matrix $\theta$ are learnable. In each iteration, $\delta_v$ is updated by maximizing the $\mathcal{L}_{visual}$ to find the worst-case visual embedding, while $\theta$ is updated by minimizing the $\mathcal{L}_{text}$ to craft a cross-image adversarial prompt. The two perturbations are optimized with opposing objectives.
} 
    \label{fig:pipeline}
\end{figure*}

\section{\ourmethod}
\label{ourmethod}
\subsection{Overview} 
In this section, we present \ourmethod, a framework for learning a cross-image transferable adversarial prompt for VLMs. The key idea of \ourmethod is to jointly optimize the adversarial suffix and image perturbation, so that the learned suffix remains effective across diverse unseen images rather than overfitting to a single image input. We first formulate the attack objective from a robust min-max perspective in the \textit{Problem Formulation} part (Sec.~\ref{sec:problem_formulation}). We then describe the overall \textit{Alternating Optimization Procedure} (Sec.~\ref{sec:alter}), where the image perturbation and adversarial suffix are updated in turn. After that, we introduce the \textit{Image Maximization Stage} (Sec.~\ref{sec:ims}), which constructs ``worst-case images'' for the current suffix. Finally, we detail the \textit{Text Minimization Stage} (Sec.~\ref{sec:lossfuc}), where the suffix is optimized using our introduced guided adversarial, coherence, and semantic alignment losses.

\subsection{Problem Formulation}
\label{sec:problem_formulation}
Let $f$ denote a target VLM. Given an image $x_v$ and a prompt template $x_t$, our goal is to optimize an adversarial suffix $\delta_t$ such that the composed prompt $x_t + \delta_t$ steers the model toward an attacker-specified target response $y_t$ across diverse unseen images. Existing prompt-based attacks typically optimize
\begin{equation}
    \min_{\delta_t} \mathcal{L}\big(f(x_v, x_t + \delta_t), y_t\big)
\end{equation}
for a fixed image-prompt pair. While effective for that specific visual context, such optimization can entangle the learned suffix with the image used during training, thereby limiting transferability to new images. To improve cross-image transferability, we move from instance-specific optimization to a distributional objective over visual inputs~\citep{wang2024trojanrobot,li2026privacy}. Let $\mathcal{D}$ denote the image distribution associated with the target template. Ideally, we would like to solve
\begin{equation}
    \min_{\delta_t} \; \mathbb{E}_{x_v \sim \mathcal{D}}
    \big[\mathcal{L}(f(x_v, x_t + \delta_t), y_t)\big].
\end{equation}
However, directly optimizing over the full visual distribution is intractable. Formally, as illustrated in Fig.~\ref{fig:pipeline}, we therefore adopt a robust min-max surrogate:
\begin{equation}
    \min_{\delta_t} \; \mathbb{E}_{x_v \sim \mathcal{D}}
    \Big[\max_{\delta_v} \mathcal{L}\big(f(x_v + \delta_v, x_t + \delta_t), y_t\big)\Big],
    \label{eq:1}
\end{equation}
where $\delta_v$ denotes image perturbations used to construct hard visual conditions for the current suffix. Intuitively, the inner maximization identifies visual inputs under which the current suffix is least effective, while the outer minimization updates the suffix so that the target response remains likely even under these difficult conditions. In this way, the learned suffix is encouraged to rely less on image-specific cues and to generalize better across unseen images.

\subsection{Alternating Optimization Procedure}
\label{sec:alter}
Directly solving the saddle-point problem in Eq.~\ref{eq:1} is difficult. We therefore adopt an alternating optimization strategy that updates the image perturbation and the text suffix in turn. In the inner loop, given the current suffix, we optimize $\delta_v$ by projected gradient ascent to construct ``\textit{worst-case images}'' for the current prompt. This step exposes the suffix to hard visual conditions during training, reducing its tendency to overfit to a fixed image.

In the outer loop, given the ``\textit{worst-case image}'', we update the adversarial suffix to increase the likelihood of the target response. Since the suffix is discrete, direct backpropagation through token identities is not possible. To address this, we introduce a continuous relaxation. Specifically, instead of optimizing the discrete suffix $\delta_t$ directly, we optimize a learnable token distribution matrix $\theta \in \mathbb{R}^{L \times V}$, where $L$ is the suffix length and $V$ is the vocabulary size. Each row $\theta_i$ represents a soft distribution over candidate tokens at position $i$. During optimization, we sample differentiable soft token vectors via Gumbel-Softmax~\citep{jang2017categorical}. Let $G_{i,j} \sim \text{Gumbel}(0,1)$ and $\tau > 0$ denote the temperature. For the $i$-th position, we compute
\begin{equation}
    (w_i)_j = 
    \frac{\exp((\theta_{i,j} + G_{i,j})/\tau)}
    {\sum_{k=1}^{V}\exp((\theta_{i,k} + G_{i,k})/\tau)},\quad \tau > 0.
    \label{eq:gbsoft}
\end{equation}
This yields a differentiable approximation to a one-hot token selection. Let $\{e^{(j)}\}_{j=1}^{V}$ be the token embeddings in the vocabulary. We then form the continuous embedding at position $i$ as
\begin{equation}
    \tilde{e}(\theta_i) = \sum_{j=1}^{V} (w_i)_j e^{(j)}, 
    \qquad \tilde{e}(\theta_i) \in \mathbb{R}^{d},
\end{equation}
where $d$ is the embedding dimension. Stacking $\{\tilde{e}(\theta_i)\}_{i=1}^{L}$ row-wise gives the continuous suffix embedding $E(\theta)$, which is concatenated with the prompt template embedding $E(x_t)$ to form $E_{\text{full}}(\theta) = [E(x_t)\, \| \, E(\theta)]$. Because this relaxation is differentiable with respect to $\theta$, gradients from the full training objective can be propagated end-to-end. After optimization, we recover the final suffix $\delta_t$ by taking the highest probability token at each position.

Next, we break down Eq.~(\ref{eq:1}) by detailing the loss functions for each of the two optimization phases, \textit{i.e.,} the image maximization stage (Sec.~\ref{sec:ims}) and the text minimization stage (Sec.~\ref{sec:lossfuc}).

\subsection{Image Maximization Stage} 
\label{sec:ims}
In this stage, we seek to identify the ``\textit{worst-case}'' image perturbation $\delta_v$ that maximally disrupts the model's adherence to the target response. This serves as a regularizer, preventing the adversarial prompt from overfitting to specific visual features. We formulate this as a maximization problem over the visual adversarial objectives $\mathcal{L}_{\text{visual}}$:
\begin{equation}
    \mathcal{L}_{\text{visual}} = - \sum_{k=1}^{|y_t|} \log p(y_k \mid E(x_v + \delta_v), E(x_t + \delta_t)),
\end{equation}
where $y_k$ is the $k$-th token of the target response $\boldsymbol{y}_{t}$.
By applying \textit{Projected Gradient Ascent} (PGA)~\citep{madry2018towards} on $\delta_v$ to maximize $\mathcal{L}_{\text{visual}}$, we push the image $x_v$ towards the decision boundary where the current adversarial prompt is least effective. Optimizing the prompt against these ``hard'' visual examples encourages the suffix to depend less on cues specific to the training images.

\begin{algorithm}[t]
\footnotesize
\caption{Optimization process of \ourmethod}
\label{algo}
\begin{algorithmic}[1]
\REQUIRE VLM $f$, target response $y_t$, clean image $x_v$, prompt template $x_t$,  
\quad image step size $\alpha_1$, prompt step size $\alpha_2$,  
\quad iteration count $K$, update interval $N$, suffix length $L$, vocabulary size $V$.
\ENSURE Textual adversarial suffix $\delta_t$

\STATE Initialize perturbed image ${x'}_v \leftarrow x_v$ \hfill \textcolor{blue}{//  Start with clean image} 
\STATE Initialize token distribution matrix $\theta\in\mathbb{R}^{L\times V}$ uniformly

\FOR{$k=1$ \TO $K$}
  \STATE {\bf Phase 1: Text Minimization} \hfill \textcolor{blue}{// Optimize suffix under current image}

    \FOR{$i=1$ \TO $L$}
      \STATE Sample $G_{i,j}\!\sim\!\mathrm{Gumbel}(0,1)$ for $j=1,\dots,V$
      \STATE Compute soft weights $(w_i)_j$ via Eq.~(\ref{eq:gbsoft})
      \STATE Calculate continuous embedding $\tilde{e}(\theta_i) \leftarrow \sum_{j=1}^{V} (w_i)_j e^{(j)}$
    \ENDFOR

    \STATE Form combined text embedding $E_{\mathrm{full}}(\theta) \leftarrow [\,E(x_t) \parallel E(\theta)]$
    \STATE Compute $\mathcal{L}_{\mathrm{text}}$ based on Eq.~(\ref{text loss})
    \STATE Update $\;\theta \leftarrow \theta - \alpha_2\,\mathrm{sign}(\nabla_{\theta}\,\mathcal{L}_{\mathrm{text}})$ \hfill \textcolor{blue}{// Gradient descent on $\theta$}
    
    \IF{$k \bmod N = 0$}
      \STATE {\bf Phase 2: Image Maximization} \hfill  \textcolor{blue}{// Construct worst-case image}
      \STATE Compute visual adversarial loss $\mathcal{L}_{\mathrm{visual}}$
      \STATE Compute gradient $g_x \leftarrow \nabla_{x'_v}\,\mathcal{L}_{\mathrm{visual}}$
      \STATE Update $\;x'_v \leftarrow x'_v + \alpha_1\,\mathrm{sign}(g_x)$ \hfill \textcolor{blue}{// Gradient ascent on $x_v$}
  \ENDIF
\ENDFOR

\STATE \textbf{Extract} discrete suffix: $\delta_{t,i} = \arg\max_j \theta_{i,j},\;\forall i=1,\dots,L$
\RETURN $\delta_t$
\end{algorithmic}
\end{algorithm}

\subsection{Text Minimization Stage} 
\label{sec:lossfuc}
Similarly, in the text optimization phase, we fix the ``\textit{worst-case images}'' derived from the image maximization stage and optimize the token distribution matrix $\theta$ to minimize the textual adversarial objectives $\mathcal{L}_{\text{text}}$. We design $\mathcal{L}_{\text{text}}$ as a composite loss that balances attack effectiveness ($\mathcal{L}_{\text{ga}}$), semantic stealthiness ($\mathcal{L}_{\text{align}}$), and linguistic fluency ($\mathcal{L}_{\text{tc}}$):
\begin{equation}
    \mathcal{L}_{\text{text}} = \mathcal{L}_{\text{ga}} + \lambda_{\text{align}}\mathcal{L}_{\text{align}} + \lambda_{\text{tc}} \mathcal{L}_{\text{tc}},
    \label{text loss}
\end{equation}
where $\lambda_{\text{align}}$ and $\lambda_{\text{tc}}$ are hyperparameters that control the contribution of the text coherence loss and semantic alignment loss.

\noindent\textbf{Guided Adversarial Loss ($\mathcal{L}_{\text{ga}}$). }
To enforce the generation of the attacker's desired response $y_t$, we minimize the negative log-likelihood of the target and define the guided adversarial loss as:
\begin{equation}
    \mathcal{L}_{\text{ga}}=\sum_{i=1}^{|y_t|}- \log (p(y_i \mid E(x_v + \delta_v), E_{full}(\theta))).
\end{equation}
Here, a lower value for $\mathcal{L}_{\text{ga}}$ indicates that the VLM's output, when guided by $\theta$, is more likely to match the target response.

\begin{table*}[t]
\centering
\small
\caption{ASRs tested on four models with different target texts. The mean and standard deviations of the ASRs are shown in the table. The best performance values for each case are highlighted in bold.}
\adjustbox{max totalsize={\textwidth}{\textheight}}{
\begin{tabular}{c|c|cccc|cccc}
\toprule[0.5pt]

\hline
\multirow{3}{*}{\textbf{Target Prompt$\downarrow$}} & \multirow{3}{*}{\textbf{Method$\downarrow$}} & \multicolumn{4}{c|}{\cellcolor[HTML]{FFF7F0}\footnotesize \textbf{MS-COCO}~\citep{lin2014microsoft}} & \multicolumn{4}{c}{\cellcolor[HTML]{FFF7F0}\textbf{ImageNet}~\citep{russakovsky2015imagenet}} \\
\cline{3-10}
 & & 
 \footnotesize \begin{tabular}[c]{@{}c@{}}  MiniGPT-4 \\ \citep{zhu2024minigpt} \end{tabular} &
 \footnotesize \begin{tabular}[c]{@{}c@{}} BLIP-2 \\ \citep{li2023blip2} \end{tabular} &
 \footnotesize \begin{tabular}[c]{@{}c@{}} InstructBLIP \\ \citep{dai2023instructblip} \end{tabular} &
 \footnotesize \begin{tabular}[c]{@{}c@{}} LLaVA-v1.5-7b \\ \citep{liu2023visual} \end{tabular} &
 \footnotesize \begin{tabular}[c]{@{}c@{}} MiniGPT-4 \\ \citep{zhu2024minigpt} \end{tabular} &
 \footnotesize \begin{tabular}[c]{@{}c@{}} BLIP-2 \\ \citep{li2023blip2} \end{tabular} &
 \footnotesize \begin{tabular}[c]{@{}c@{}} InstructBLIP \\ \citep{dai2023instructblip} \end{tabular} &

 \footnotesize \begin{tabular}[c]{@{}c@{}} LLaVA-v1.5-7b \\ \citep{liu2023visual} \end{tabular} \\
\hline

 \multirow{9}{*}{ {\textit{``Yes''}}}

  & {\ct{Multi-I} (Self-Constructed)}
 & 0.36\uncertsub{6.38\text{e-3}}
 & 0.31\uncertsub{6.34\text{e-3}}
 & 0.28\uncertsub{1.16\text{e-2}}
 & 0.25\uncertsub{5.65\text{e-2}}
 
 & 0.36\uncertsub{3.74\text{e-3}}
 & 0.32\uncertsub{3.77\text{e-3}}
 & 0.26\uncertsub{4.92\text{e-3}}
 & 0.24\uncertsub{8.91\text{e-3}}
 \\

 & {\ct{PGD-BERT}~\citep{waghela2024enhancing}}
 & 0.30\uncertsub{2.12\text{e-2}}
 & 0.27\uncertsub{4.17\text{e-3}}
 & 0.28\uncertsub{2.34\text{e-2}}
 & 0.28\uncertsub{2.16\text{e-2}}
 
 & 0.31\uncertsub{2.66\text{e-3}}
 & 0.29\uncertsub{1.17\text{e-3}}
 & 0.27\uncertsub{3.41\text{e-2}}
 & 0.28\uncertsub{1.16\text{e-2}}
 \\

  & {\ct{BAP}~\citep{ying2024jailbreak}}
 & 0.34\uncertsub{2.21\text{e-3}}
 & 0.32\uncertsub{4.33\text{e-2}}
 & 0.33\uncertsub{5.21\text{e-2}}
 & 0.28\uncertsub{4.93\text{e-3}}
 
 & 0.32\uncertsub{2.07\text{e-3}} 
 & 0.30\uncertsub{2.21\text{e-2}}
 & 0.31\uncertsub{3.56\text{e-2}}
 & 0.30\uncertsub{1.02\text{e-2}}
 
 \\

 & {\ct{GCG-Transfer}~\citep{zou2023universal}}
 & 0.40\uncertsub{2.16\text{e-3}}
 & 0.42\uncertsub{1.47\text{e-2}}
 & 0.38\uncertsub{1.25\text{e-3}}
 & 0.30\uncertsub{3.71\text{e-3}} 
 
 & 0.37\uncertsub{2.12\text{e-2}}
 & 0.36\uncertsub{5.72\text{e-3}}
 & 0.36\uncertsub{1.70\text{e-3}}
 & 0.29\uncertsub{8.10\text{e-3}} 
  \\
 
& {\ct{PAIR}~\citep{chao2023jailbreaking}}
 & 0.38\uncertsub{2.55\text{e-2}}
 & 0.38\uncertsub{1.26\text{e-2}}
 & 0.35\uncertsub{1.87\text{e-2}}
  & 0.31\uncertsub{2.36\text{e-2}}
  
  & 0.37\uncertsub{2.82\text{e-2}}
 & 0.38\uncertsub{8.65\text{e-3}}
 & 0.35\uncertsub{2.02\text{e-2}}
 & 0.28\uncertsub{5.97\text{e-2}}
 \\

 & {\ct{TAP}~\citep{mehrotra2023treeOfAttacks}}
 & 0.43\uncertsub{1.31\text{e-2}}
 & 0.39\uncertsub{2.30\text{e-2}}
 & 0.35\uncertsub{6.80\text{e-3}}
 & 0.33\uncertsub{3.94\text{e-2}}
 
 & 0.47\uncertsub{1.92\text{e-2}} 
 & 0.46\uncertsub{1.36\text{e-2}}
 & 0.32\uncertsub{1.70\text{e-2}}
 & 0.30\uncertsub{5.11\text{e-3}}
 
 \\

 & {\ct{AutoDAN}~\citep{liu2024autodan}}
 & 0.37\uncertsub{6.18\text{e-3}}
 & 0.33\uncertsub{1.07\text{e-2}}
 & 0.29\uncertsub{1.11\text{e-2}}
 & 0.25\uncertsub{3.11\text{e-2}}
 
 & 0.35\uncertsub{9.20\text{e-3}}
 & 0.32\uncertsub{8.65\text{e-3}}
 & 0.27\uncertsub{8.81\text{e-3}}
 & 0.26\uncertsub{4.56\text{e-2}}
\\

 & {\ct{DeGCG}~\citep{liu2024advancing}}
 & 0.41\uncertsub{6.18\text{e-3}}
 & 0.37\uncertsub{7.98\text{e-3}}
 & 0.35\uncertsub{3.84\text{e-2}}
 & 0.31\uncertsub{2.42\text{e-2}}

 & 0.42\uncertsub{5.67\text{e-2}}
 & 0.40\uncertsub{2.12\text{e-2}}
 & 0.34\uncertsub{7.12\text{e-3}}
 & 0.27\uncertsub{8.09\text{e-3}}

\\

 &  { \cellcolor{gray!8} \ourmethod \textbf{(Ours)}}
 &   \cellcolor{gray!8} ${\hspace*{-0.3em}\textbf{0.64\uncertsub{\textbf{3.15e-2}}}}$
 &   \cellcolor{gray!8} $\hspace*{-0.3em}\textbf{0.60\uncertsub{\textbf{5.80e-3}}}$
 &   \cellcolor{gray!8} $\hspace*{-0.3em}\textbf{0.55\uncertsub{\textbf{6.94e-3}}}$
 &   \cellcolor{gray!8} $\hspace*{-0.3em}\textbf{0.36\uncertsub{\textbf{1.93e-2}}}$
 
 &  \cellcolor{gray!8}  $\hspace*{-0.3em}\textbf{0.63\uncertsub{\textbf{2.72e-2}}}$ 
 &   \cellcolor{gray!8} $\hspace*{-0.3em}\textbf{0.61\uncertsub{\textbf{1.16e-2}}}$
 &  \cellcolor{gray!8}  $\hspace*{-0.3em}\textbf{0.54\uncertsub{\textbf{3.69e-2}}}$
&   \cellcolor{gray!8} $\hspace*{-0.3em}\textbf{0.34\uncertsub{\textbf{1.77e-2}}}$
 \\
 \hline
 
\multirow{9}{*}{ {\textit{``too late''}}}
  & {\ct{Multi-I} (Self-Constructed)}
 & 0.39\uncertsub{9.43\text{e-3}}
 & 0.37\uncertsub{1.63\text{e-2}}
 & 0.32\uncertsub{4.71\text{e-3}}
 & 0.30\uncertsub{9.10\text{e-3}}
 
 & 0.40\uncertsub{1.25\text{e-2}}
 & 0.36\uncertsub{8.16\text{e-3}}
 & 0.27\uncertsub{1.58\text{e-2}}
 & 0.28\uncertsub{4.02\text{e-2}}
  \\

  & {\ct{PGD-BERT}~\citep{waghela2024enhancing}}
 & 0.30\uncertsub{2.82\text{e-3}}
 & 0.27\uncertsub{3.42\text{e-3}}
 & 0.24\uncertsub{2.54\text{e-2}}
 & 0.21\uncertsub{1.80\text{e-2}}
 
 & 0.27\uncertsub{5.13\text{e-3}}
 & 0.25\uncertsub{3.62\text{e-2}}
 & 0.26\uncertsub{2.53\text{e-3}}
 & 0.20\uncertsub{3.19\text{e-2}}
   \\

 & {\ct{BAP}~\citep{ying2024jailbreak}}
 & 0.32\uncertsub{5.03\text{e-2}}
 & 0.30\uncertsub{7.10\text{e-3}}
 & 0.33\uncertsub{2.81\text{e-3}}
 & 0.28\uncertsub{3.63\text{e-3}}
 
 & 0.34\uncertsub{4.74\text{e-3}} 
 & 0.35\uncertsub{5.35\text{e-2}}
 & 0.32\uncertsub{5.24\text{e-2}}
 & 0.27\uncertsub{1.34\text{e-2}}
 
 \\
  
 & {\ct{GCG-Transfer}~\citep{zou2023universal}}
 & 0.39\uncertsub{6.60\text{e-3}}
 & 0.29\uncertsub{2.03\text{e-2}}
 & 0.27\uncertsub{2.42\text{e-2}}
 & 0.23\uncertsub{5.38\text{e-3}}
 
 & 0.37\uncertsub{2.41\text{e-2}}
 & 0.30\uncertsub{6.38\text{e-3}}
 & 0.27\uncertsub{1.55\text{e-2}}
 & 0.24\uncertsub{6.63\text{e-3}}
 \\
 
 &  {\ct{PAIR}~\citep{chao2023jailbreaking}}
 & 0.40\uncertsub{2.32\text{e-2}}
 & 0.34\uncertsub{1.32\text{e-2}}
 & 0.32\uncertsub{1.53\text{e-2}}
 & 0.27\uncertsub{2.37\text{e-2}}
 
 & 0.38\uncertsub{3.48\text{e-2}}
 & 0.34\uncertsub{1.47\text{e-3}}
 & 0.33\uncertsub{1.82\text{e-2}}
 & 0.29\uncertsub{3.18\text{e-2}}
 
 \\

 & {\ct{TAP}~\citep{mehrotra2023treeOfAttacks}}
 & 0.42\uncertsub{1.27\text{e-2}}
 & 0.37\uncertsub{1.52\text{e-2}}
 & 0.36\uncertsub{4.50\text{e-3}}
 & 0.26\uncertsub{2.64\text{e-2}}
 
 & 0.41\uncertsub{1.28\text{e-2}} 
 & 0.38\uncertsub{1.25\text{e-3}}
 & 0.33\uncertsub{1.05\text{e-2}}
 & 0.27\uncertsub{8.62\text{e-3}}
 \\

 & {\ct{AutoDAN}~\citep{liu2024autodan}}
 & 0.35\uncertsub{1.82\text{e-2}}
 & 0.31\uncertsub{7.79\text{e-3}}
 & 0.27\uncertsub{8.99\text{e-3}}
 & 0.25\uncertsub{5.79\text{e-3}}
 
 & 0.34\uncertsub{2.87\text{e-3}}
  & 0.31\uncertsub{5.72\text{e-3}}
 & 0.29\uncertsub{3.27\text{e-3}}
 & 0.26\uncertsub{9.21\text{e-3}}
 \\

 & {\ct{DeGCG}~\citep{liu2024advancing}}
 & 0.36\uncertsub{6.22\text{e-3}}
 & 0.34\uncertsub{2.05\text{e-3}}
 & 0.31\uncertsub{1.16\text{e-2}}
 & 0.26\uncertsub{2.31\text{e-2}}
 
 & 0.35\uncertsub{1.40\text{e-2}}
 & 0.31\uncertsub{5.42\text{e-3}}
 & 0.29\uncertsub{9.21\text{e-3}}
 & 0.26\uncertsub{6.55\text{e-3}}
 \\
 
 &  \cellcolor{gray!8}  {\ourmethod \textbf{(Ours)}}
 &  \cellcolor{gray!8}  $\hspace*{-0.3em}\textbf{0.58\uncertsub{\textbf{6.16e-3}}}$
 &  \cellcolor{gray!8}  $\hspace*{-0.3em}\textbf{0.55\uncertsub{\textbf{8.65e-3}}}$
 &  \cellcolor{gray!8}  $\hspace*{-0.3em}\textbf{0.51\uncertsub{\textbf{2.73e-2}}}$
 &  \cellcolor{gray!8}  $\hspace*{-0.3em}\textbf{0.34\uncertsub{\textbf{1.59e-2}}}$
 
 &  \cellcolor{gray!8}  $\hspace*{-0.3em}\textbf{0.59\uncertsub{\textbf{1.09e-2}}}$
 &  \cellcolor{gray!8}  $\hspace*{-0.3em}\textbf{0.56\uncertsub{\textbf{1.11e-2}}}$
 &  \cellcolor{gray!8}  $\hspace*{-0.3em}\textbf{0.47\uncertsub{\textbf{1.49e-2}}}$
 &  \cellcolor{gray!8}  $\hspace*{-0.3em}\textbf{0.35\uncertsub{\textbf{4.13e-3}}}$
 \\
\hline

\multirow{9}{*}{ {\textit{``I'm sorry''}}}

 & {\ct{Multi-I} (Self-Constructed)}
 & 0.43\uncertsub{1.89\text{e-2}}
 & 0.40\uncertsub{1.57\text{e-2}}
 & 0.35\uncertsub{1.79\text{e-2}}
 & 0.28\uncertsub{9.24\text{e-3}}
 
 & 0.38\uncertsub{1.34\text{e-2}}
 & 0.37\uncertsub{1.25\text{e-3}}
 & 0.35\uncertsub{7.12\text{e-3}}
 & 0.29\uncertsub{8.29\text{e-3}}
\\

 & {\ct{PGD-BERT}~\citep{waghela2024enhancing}}
 & 0.33\uncertsub{3.92\text{e-2}}
 & 0.32\uncertsub{5.64\text{e-3}}
 & 0.33\uncertsub{2.83\text{e-2}}
 & 0.23\uncertsub{1.94\text{e-2}}
 
 & 0.36\uncertsub{1.27\text{e-3}}
  & 0.31\uncertsub{5.47\text{e-2}}
 & 0.29\uncertsub{2.75\text{e-3}}
 & 0.21\uncertsub{3.84\text{e-2}}
 \\

 & {\ct{BAP}~\citep{ying2024jailbreak}}
 & 0.34\uncertsub{4.02\text{e-2}}
 & 0.32\uncertsub{1.04\text{e-2}}
 & 0.33\uncertsub{2.12\text{e-3}}
 & 0.25\uncertsub{2.47\text{e-2}}
 
 & 0.32\uncertsub{3.47\text{e-2}} 
 & 0.30\uncertsub{4.26\text{e-3}}
 & 0.31\uncertsub{5.73\text{e-2}}
 & 0.23\uncertsub{6.04\text{e-3}}
 
 \\

  & {\ct{GCG-Transfer}~\citep{zou2023universal}}
 & 0.35\uncertsub{8.34\text{e-3}}
 & 0.34\uncertsub{9.09\text{e-3}}
 & 0.29\uncertsub{1.63\text{e-2}}
 & 0.25\uncertsub{4.00\text{e-3}}
 
 & 0.38\uncertsub{1.62\text{e-2}}
  & 0.31\uncertsub{1.43\text{e-2}}
 & 0.29\uncertsub{1.60\text{e-2}}
 & 0.24\uncertsub{5.10\text{e-3}}
 
\\

 & {\ct{PAIR}~\citep{chao2023jailbreaking}}
 & 0.41\uncertsub{2.20\text{e-2}}
 & 0.36\uncertsub{1.68\text{e-2}}
 & 0.33\uncertsub{1.95\text{e-2}}
 & 0.25\uncertsub{5.42\text{e-2}}
 
 & 0.43\uncertsub{1.04\text{e-2}}
 & 0.34\uncertsub{9.50\text{e-3}}
 & 0.33\uncertsub{1.10\text{e-2}}
 & 0.27\uncertsub{4.37\text{e-3}}
 
 \\
 & {\ct{TAP}~\citep{mehrotra2023treeOfAttacks}}
 & 0.46\uncertsub{2.94\text{e-2}}
 & 0.41\uncertsub{1.05\text{e-2}}
 & 0.34\uncertsub{1.03\text{e-2}}
  & 0.27\uncertsub{5.72\text{e-2}}
 
 & 0.45\uncertsub{2.59\text{e-3}}
 & 0.39\uncertsub{5.23\text{e-3}}
 & 0.35\uncertsub{1.85\text{e-2}}
 & 0.26\uncertsub{1.08\text{e-2}}
 
 \\

 & {\ct{AutoDAN}~\citep{liu2024autodan}}
 & 0.45\uncertsub{1.84\text{e-2}}
 & 0.37\uncertsub{1.24\text{e-2}}
 & 0.33\uncertsub{5.31\text{e-3}}
 & 0.25\uncertsub{8.22\text{e-3}}
 
 & 0.39\uncertsub{5.56\text{e-3}}
 & 0.32\uncertsub{5.89\text{e-3}}
 & 0.34\uncertsub{7.48\text{e-3}}
 & 0.24\uncertsub{3.70\text{e-3}}
 
 \\

 & {\ct{DeGCG}~\citep{liu2024advancing}}
 & 0.42\uncertsub{1.72\text{e-2}}
 & 0.37\uncertsub{2.25\text{e-2}}
 & 0.32\uncertsub{1.56\text{e-2}}
 & 0.27\uncertsub{1.90\text{e-2}}
 
 & 0.42\uncertsub{5.10\text{e-3}}
 & 0.37\uncertsub{8.65\text{e-3}}
 & 0.35\uncertsub{8.38\text{e-3}}
 & 0.24\uncertsub{1.26\text{e-2}}
 
\\

 &  \cellcolor{gray!8}  {\ourmethod \textbf{(Ours)}}
 &  \cellcolor{gray!8}  $\hspace*{-0.2em}\textbf{0.62\uncertsub{\textbf{1.32e-2}}}$
 &  \cellcolor{gray!8}  $\hspace*{-0.2em}\textbf{0.60\uncertsub{\textbf{7.32e-3}}}$
 &  \cellcolor{gray!8}  $\hspace*{-0.2em}\textbf{0.55\uncertsub{\textbf{1.04e-2}}}$
 &  \cellcolor{gray!8}  $\hspace*{-0.2em}\textbf{0.32\uncertsub{\textbf{5.29e-3}}}$
 
 &  \cellcolor{gray!8}  $\hspace*{-0.2em}\textbf{0.62\uncertsub{\textbf{1.52e-2}}} $
 &  \cellcolor{gray!8}  $\hspace*{-0.2em}\textbf{0.61\uncertsub{\textbf{2.02e-2}}} $
 &  \cellcolor{gray!8}  $\hspace*{-0.2em}\textbf{0.54\uncertsub{\textbf{1.78e-2}}} $
 &  \cellcolor{gray!8}  $\hspace*{-0.2em}\textbf{0.33\uncertsub{\textbf{2.07e-2}}}$
 
 \\

\hline

\multirow{9}{*}{ {\textit{``This image is in black''}}}

 & {\ct{Multi-I} (Self-Constructed)}
 & 0.37\uncertsub{2.86\text{e-2}}
 & 0.34\uncertsub{5.36\text{e-2}}
 & 0.35\uncertsub{6.25\text{e-3}}
 & 0.31\uncertsub{2.68\text{e-2}}
 
 & 0.42\uncertsub{5.53\text{e-2}}
 & 0.41\uncertsub{3.62\text{e-2}}
 & 0.37\uncertsub{2.85\text{e-2}}
 & 0.25\uncertsub{5.38\text{e-2}}
 
 \\
 & {\ct{PGD-BERT}~\citep{waghela2024enhancing}}
 & 0.29\uncertsub{1.36\text{e-2}}
 & 0.26\uncertsub{2.38\text{e-2}}
 & 0.23\uncertsub{3.84\text{e-3}}
 & 0.20\uncertsub{1.37\text{e-2}}
 
 & 0.25\uncertsub{2.41\text{e-2}}
 & 0.24\uncertsub{5.21\text{e-2}}
 & 0.24\uncertsub{9.35\text{e-3}}
 & 0.19\uncertsub{7.14\text{e-3}}

 \\
 
 & {\ct{BAP}~\citep{ying2024jailbreak}}
 & 0.30\uncertsub{2.08\text{e-2}}
 & 0.32\uncertsub{9.17\text{e-3}}
 & 0.31\uncertsub{1.86\text{e-2}}
 & 0.28\uncertsub{2.57\text{e-2}}
 
 & 0.39\uncertsub{5.68\text{e-3}} 
 & 0.32\uncertsub{2.63\text{e-3}}
 & 0.36\uncertsub{4.13\text{e-3}}
 & 0.27\uncertsub{1.34\text{e-2}}
 
 \\

 & {\ct{GCG-Transfer}~\citep{zou2023universal}}
 & 0.44\uncertsub{4.56\text{e-2}}
 & 0.43\uncertsub{5.49\text{e-3}}
 & 0.40\uncertsub{1.38\text{e-2}}
 & 0.24\uncertsub{6.62\text{e-3}}
 
 & 0.41\uncertsub{2.41\text{e-2}}
 & 0.41\uncertsub{3.38\text{e-3}}
 & 0.36\uncertsub{4.62\text{e-2}}
 & 0.24\uncertsub{1.54\text{e-2}}
 \\ 
 
 &  {\ct{PAIR}~\citep{chao2023jailbreaking}}
 & 0.41\uncertsub{3.32\text{e-2}}
 & 0.37\uncertsub{2.32\text{e-2}}
 & 0.36\uncertsub{2.63\text{e-2}}
 & 0.27\uncertsub{4.81\text{e-2}}
 
 & 0.40\uncertsub{3.48\text{e-2}}
 & 0.33\uncertsub{1.47\text{e-3}}
 & 0.35\uncertsub{1.82\text{e-2}}
 & 0.28\uncertsub{3.18\text{e-2}}
 
 \\

 & {\ct{TAP}~\citep{mehrotra2023treeOfAttacks}}
 & 0.45\uncertsub{2.76\text{e-2}}
 & 0.43\uncertsub{6.54\text{e-3}}
 & 0.40\uncertsub{3.52\text{e-2}}
 & 0.25\uncertsub{5.24\text{e-3}}
 
 & 0.43\uncertsub{2.34\text{e-2}} 
 & 0.40\uncertsub{4.36\text{e-2}}
 & 0.34\uncertsub{2.95\text{e-3}}
 & 0.27\uncertsub{8.62\text{e-2}}
 \\
 
 & {\ct{AutoDAN}~\citep{liu2024autodan}}
 & 0.37\uncertsub{5.82\text{e-3}}
 & 0.32\uncertsub{4.26\text{e-2}}
 & 0.32\uncertsub{2.49\text{e-3}}
 & 0.26\uncertsub{3.75\text{e-2}}
 
 & 0.38\uncertsub{6.14\text{e-3}}
  & 0.40\uncertsub{3.27\text{e-3}}
 & 0.37\uncertsub{6.22\text{e-2}}
 & 0.24\uncertsub{1.36\text{e-2}}
 
\\

 & {\ct{DeGCG}~\citep{liu2024advancing}}
 & 0.38\uncertsub{5.12\text{e-3}}
 & 0.31\uncertsub{2.05\text{e-3}}
 & 0.32\uncertsub{2.52\text{e-2}}
 & 0.22\uncertsub{5.11\text{e-2}}
 
 & 0.35\uncertsub{1.40\text{e-2}}
 & 0.33\uncertsub{2.55\text{e-2}}
 & 0.26\uncertsub{4.57\text{e-2}}
 & 0.24\uncertsub{2.35\text{e-2}}
 
\\

 &  \cellcolor{gray!8}  {\ourmethod \textbf{(Ours)}}
 &  \cellcolor{gray!8}  $\hspace*{-0.3em}\textbf{0.57\uncertsub{\textbf{6.16e-3}}}$
 &  \cellcolor{gray!8}  $\hspace*{-0.3em}\textbf{0.54\uncertsub{\textbf{8.65e-3}}}$
 &  \cellcolor{gray!8}  $\hspace*{-0.3em}\textbf{0.52\uncertsub{\textbf{2.73e-2}}}$
 &  \cellcolor{gray!8}  $\hspace*{-0.3em}\textbf{0.35\uncertsub{\textbf{1.59e-2}}}$
 &  \cellcolor{gray!8}  $\hspace*{-0.3em}\textbf{0.58\uncertsub{\textbf{1.03e-2}}} $
 &  \cellcolor{gray!8}  $\hspace*{-0.3em}\textbf{0.55\uncertsub{\textbf{1.16e-2}}} $
 &  \cellcolor{gray!8}  $\hspace*{-0.3em}\textbf{0.53\uncertsub{\textbf{1.99e-2}}} $
 &  \cellcolor{gray!8}  $\hspace*{-0.3em}\textbf{0.36\uncertsub{\textbf{4.33e-3}}}$
 \\
\hline

\multirow{9}{*}{ {\textit{``Absolutely! I can help you with that''}}}

 & {\ct{Multi-I} (Self-Constructed)}
 & 0.42\uncertsub{1.32\text{e-2}}
 & 0.35\uncertsub{7.14\text{e-3}}
 & 0.33\uncertsub{1.42\text{e-2}}
 & 0.28\uncertsub{2.00\text{e-2}}
 
 & 0.32\uncertsub{2.52\text{e-2}}
 & 0.29\uncertsub{1.41\text{e-2}}
 & 0.27\uncertsub{3.51\text{e-3}}
 & 0.26\uncertsub{3.42\text{e-2}}
 
 \\

 & {\ct{PGD-BERT}~\citep{waghela2024enhancing}}
 & 0.30\uncertsub{1.62\text{e-2}}
 & 0.25\uncertsub{1.25\text{e-2}}
 & 0.23\uncertsub{3.26\text{e-2}}
 & 0.21\uncertsub{2.15\text{e-2}}
 
 & 0.23\uncertsub{4.26\text{e-3}}
 & 0.23\uncertsub{2.52\text{e-2}}
 & 0.24\uncertsub{3.54\text{e-3}}
 & 0.16\uncertsub{2.41\text{e-2}}
 
  \\

 & {\ct{BAP}~\citep{ying2024jailbreak}}
 & 0.34\uncertsub{4.03\text{e-2}}
 & 0.30\uncertsub{2.10\text{e-3}}
 & 0.30\uncertsub{5.11\text{e-3}}
 & 0.25\uncertsub{1.53\text{e-2}}
 
 & 0.35\uncertsub{6.61\text{e-3}} 
 & 0.32\uncertsub{1.36\text{e-2}}
 & 0.31\uncertsub{4.25\text{e-2}}
 & 0.25\uncertsub{7.14\text{e-3}}
 
 \\

  & {\ct{GCG-Transfer}~\citep{zou2023universal}}
 & 0.35\uncertsub{1.53\text{e-2}}
 & 0.34\uncertsub{5.53\text{e-2}}
 & 0.28\uncertsub{3.46\text{e-2}}
 & 0.24\uncertsub{2.34\text{e-2}}
 
 & 0.31\uncertsub{7.11\text{e-3}}
 & 0.30\uncertsub{7.09\text{e-3}}
 & 0.26\uncertsub{3.01\text{e-2}}
 & 0.22\uncertsub{9.03\text{e-3}}
 
 \\

 &  {\ct{PAIR}~\citep{chao2023jailbreaking}}
 & 0.34\uncertsub{1.51\text{e-2}}
 & 0.32\uncertsub{2.14\text{e-2}}
 & 0.32\uncertsub{4.52\text{e-2}}
 & 0.26\uncertsub{6.12\text{e-3}}
 
 & 0.34\uncertsub{1.47\text{e-2}}
 & 0.32\uncertsub{2.11\text{e-3}}
 & 0.32\uncertsub{4.62\text{e-2}}
 & 0.28\uncertsub{3.63\text{e-2}}
 
 \\

 & {\ct{TAP}~\citep{mehrotra2023treeOfAttacks}}
 & 0.40\uncertsub{2.36\text{e-2}}
 & 0.39\uncertsub{2.51\text{e-2}}
 & 0.34\uncertsub{5.21\text{e-3}}
 & 0.22\uncertsub{1.25\text{e-2}}
 
 & 0.40\uncertsub{9.68\text{e-2}} 
 & 0.37\uncertsub{4.43\text{e-3}}
 & 0.31\uncertsub{1.35\text{e-2}}
 & 0.26\uncertsub{4.34\text{e-3}}
 
 \\
 
 & {\ct{AutoDAN}~\citep{liu2024autodan}}
 & 0.35\uncertsub{1.15\text{e-2}}
 & 0.33\uncertsub{2.24\text{e-3}}
 & 0.28\uncertsub{4.68\text{e-3}}
 & 0.26\uncertsub{2.35\text{e-3}}
 
 & 0.35\uncertsub{2.72\text{e-3}}
  & 0.32\uncertsub{4.32\text{e-3}}
 & 0.29\uncertsub{6.17\text{e-3}}
 & 0.24\uncertsub{5.25\text{e-3}}
 
\\

 & {\ct{DeGCG}~\citep{liu2024advancing}}
 & 0.32\uncertsub{1.25\text{e-3}}
 & 0.33\uncertsub{5.43\text{e-3}}
 & 0.31\uncertsub{4.26\text{e-2}}
 & 0.23\uncertsub{8.54\text{e-2}}
 
 & 0.34\uncertsub{7.10\text{e-3}}
 & 0.32\uncertsub{2.25\text{e-2}}
 & 0.26\uncertsub{1.39\text{e-2}}
 & 0.24\uncertsub{2.46\text{e-2}}
 
\\

 &  \cellcolor{gray!8}  {\ourmethod \textbf{(Ours)}}
 &  \cellcolor{gray!8}  $\hspace*{-0.3em}\textbf{0.54\uncertsub{\textbf{7.46e-3}}}$
 &  \cellcolor{gray!8}  $\hspace*{-0.3em}\textbf{0.53\uncertsub{\textbf{1.15e-2}}}$
 &  \cellcolor{gray!8}  $\hspace*{-0.3em}\textbf{0.48\uncertsub{\textbf{3.15e-2}}}$
 &  \cellcolor{gray!8}  $\hspace*{-0.3em}\textbf{0.32\uncertsub{\textbf{5.14e-2}}}$
 &  \cellcolor{gray!8}  $\hspace*{-0.3em}\textbf{0.55\uncertsub{\textbf{3.26e-2}}}$
 &  \cellcolor{gray!8}  $\hspace*{-0.3em}\textbf{0.56\uncertsub{\textbf{2.84e-2}}}$
 &  \cellcolor{gray!8}  $\hspace*{-0.3em}\textbf{0.50\uncertsub{\textbf{9.14e-3}}}$
 &  \cellcolor{gray!8}  $\hspace*{-0.3em}\textbf{0.33\uncertsub{\textbf{2.35e-2}}}$
 \\
\bottomrule[0.5pt]
\toprule[0.5pt]
\end{tabular}
}
\label{tab:different_target}
\end{table*}

\noindent\textbf{Semantic Alignment Loss ($\mathcal{L}_{\mathrm{align}}$). }
Because our adversarial suffix is designed to steer VLMs toward attacker-specified outputs---sometimes contradicting the visual evidence---well-aligned VLMs may flag such behavior as anomalous under their safety and alignment priors~\citep{ouyang2022training, touvron2023llama2}, thereby suppressing the target response and rendering the attack ineffective. To prevent such detection and ensure the suffix is treated as a legitimate prompt, we propose a novel Semantic Alignment Loss $\mathcal{L}_{\text{align}}$ that encourages suffix embeddings to blend into the distribution of benign instructions, while pushing them away from the distribution of harmful instructions. By making the suffix appear similar to those of normal prompts, this loss encourages VLMs to generate a naturalistic and fluent reply. For instance, once the model is steered to output the initial target phrase (\textit{e.g.,} ``I'm sorry''), its standard next-token prediction mechanism is more likely to produce a plausible and contextually appropriate continuation, enhancing the overall stealthiness of the attack (see Fig.~\ref{example}). Formally,  the representations of benign instructions $\mathcal{I}_{\text{benign}}$ and harmful instructions $\mathcal{I}_{\text{harmful}}$ can be formulated as:
\begin{equation}
\begin{gathered}
E^{+} = \left\{ \mathrm{Rep}(I^{+}) \mid I^{+} \in \mathcal{I}_{\text{benign}} \right\}, \\
E^{-} = \left\{ \mathrm{Rep}(I^{-}) \mid I^{-} \in \mathcal{I}_{\text{harmful}} \right\},
\end{gathered}
\end{equation}
where $\mathrm{Rep}(I)$ is the function that takes an instruction $I$, feeds it through the VLM's text encoder, and returns the hidden state vector at the final token position. 

The $\mathcal{L}_{\mathrm{align}}$ comprises two contrastive terms. Let $\mathbf{c}^{+} = \frac{1}{|E^{+}|}\sum_{e\in E^{+}}e$ and $\mathbf{c}^{-} = \frac{1}{|E^{-}|}\sum_{e\in E^{-}}e$ denote the centroids of benign and harmful instruction embeddings. The terms are defined as:
\begin{align}
    \mathcal{L}_{\mathrm{cls}} = \big\| E_{\mathrm{full}}(\theta) - \mathbf{c}^{+} \big\|_{2}^{2},  \quad
    \mathcal{L}_{\mathrm{far}} = - \big\| E_{\mathrm{full}}(\theta) - \mathbf{c}^{-} \big\|_{2}^{2}.
\end{align}
$\mathcal{L}_{\mathrm{cls}}$ encourages the adversarial embeddings $E_{\text{full}}(\theta)$ to approach the benign centroid $\mathbf{c}^{+}$, thereby reducing the semantic distance to clean sentences to overcome the model's safety alignment. Conversely, $\mathcal{L}_{\mathrm{far}}$ ensures that the embeddings diverge from the harmful centroid $\mathbf{c}^{-}$, enhancing concealment within the alignment layer~\citep{li2024safety,gao2024shaping}.

Finally, our semantic alignment loss is formulated as:
\begin{equation}
    \mathcal{L}_{\text{align}} = {\mathcal{L}_{\text{cls}}} + \mu {\mathcal{L}_{\text{far}}},
    \label{mimicloss}
\end{equation}
where $\mu$ balances the trade-off between misalignment and the retention of benign semantics.

\noindent\textbf{Text Coherence Loss ($\mathcal{L}_{\text{tc}}$).}
To enhance the stealth and fluency of our adversarial suffix, making it less distinguishable from genuine language, we enforce that the soft token at each position aligns with the model's intrinsic next-token prediction based on the preceding context. 
Let $f_{\text{LM}}$ denote the VLM's language modeling head. At step $i$, given the prefix sequence of continuous embeddings $E_{1:i}(\theta)$, the model predicts a probability distribution over the vocabulary for the next token: $\boldsymbol{\hat p}_{i+1} = \text{Softmax}(f_{\text{LM}}(E_{1:i}(\theta)))$.
We define the text coherence loss as the cross-entropy between our learnable soft token vector $\boldsymbol{w}_{i+1}$ and $\boldsymbol{\hat p}_{i+1}$:

\begin{equation}
    \mathcal{L}_{\text{tc}}=-\sum_{i=0}^{L-1} \sum_{j=1}^{V} (\boldsymbol{w}_{i+1})_j \cdot \log (\boldsymbol{ \hat p}_{i+1})_j.
\end{equation}
By minimizing $\mathcal{L}_{\text{tc}}$, we penalize token choices that are statistically improbable under the VLM's pre-trained distribution. This effectively confines the adversarial search space to the manifold of fluent, natural text. Algorithm~\ref{algo} illustrates the complete attack process.

\section{Experiments and Results}

\subsection{Experimental Setup}
\label{Experiments Setup}

\noindent\textbf{Datasets.} 
We use prompt templates from the \textit{VQA}~\citep{goyal2017making} and \textit{GQA}~\citep{hudson2019gqa} dataset. For a rigorous assessment of \ourmethod's cross-image transferability, our evaluation leverages two diverse datasets: a subset of the MS-COCO validation set~\citep{lin2014microsoft} and ImageNet validation set~\citep{russakovsky2015imagenet}. 

\noindent\textbf{Models.}
Following SoTA adversarial attacks~\citep{Wang2024White,luo2024image,qi2024visual} on VLMs, we select four representative, open-source VLMs: (1) MiniGPT-4~\citep{zhu2024minigpt}, (2) BLIP-2~\citep{li2023blip2}, (3) InstructBLIP~\citep{dai2023instructblip}, (4) LLaVA-v1.5-7b~\citep{liu2023visual}. We adopt the open-source \textbf{Vicuna-13b} for MiniGPT-4 and InstructBLIP, \textbf{OPT-2.7b} for BLIP-2. They are widely adopted in academic and industrial research.

\noindent\textbf{Metrics.}
We use \textit{Attack Success Rate} (ASR) as the primary metric. An attack is counted as successful if the generated response matches the attacker-specified target response. We report ASR over three independent runs. 

\noindent\textbf{Competitors.}
We benchmark \ourmethod against SoTA adversarial attacks, including white-box methods (\ct{GCG-Transfer}~\citep{zou2023universal}, \ct{DeGCG}~\citep{liu2024advancing}, \ct{AutoDAN}~\citep{liu2024autodan}, \ct{PGD-BERT}~\citep{waghela2024enhancing}) and black-box methods (\ct{PAIR}~\citep{chao2023jailbreaking} and \ct{TAP}~\citep{mehrotra2023treeOfAttacks}). We also include a dual-modal baseline \ct{BAP}~\citep{ying2024jailbreak} by using only its textual adversarial component. Additionally, we implement a self-constructed baseline \ct{Multi-I} (introduced in Sec.~\ref{sec:challenes}). While these methods are prominent in various domains (\eg jailbreaking), their fundamental mechanism is a form of \textit{targeted adversarial attack} that optimizes a prompt to generate a specific string. 

\subsection{Main Results}
\label{sec:main_results}
\begin{figure}[t]
    \centering
    \includegraphics[width=0.49\textwidth]{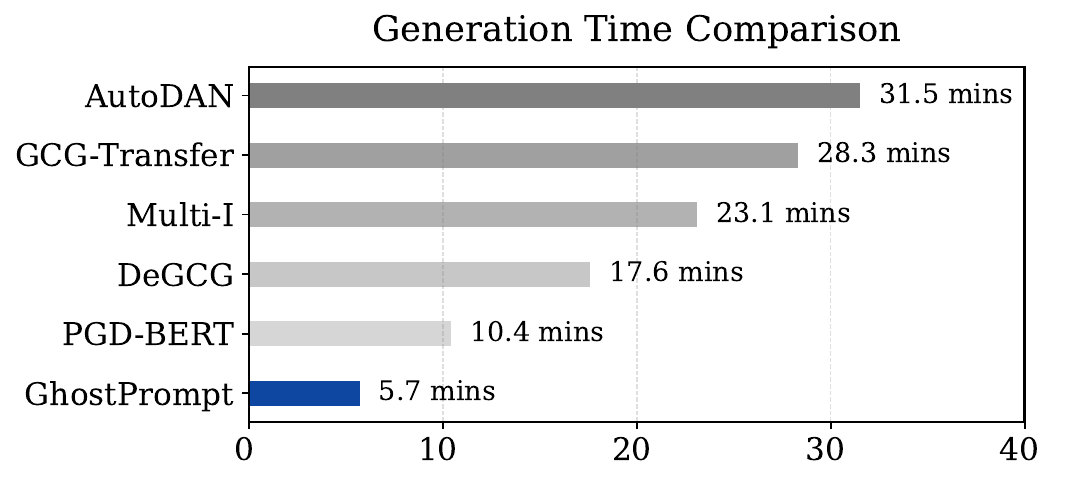}
    \caption{Average training times to generate a complete adversarial prompt across all target VLMs.}
    \label{time spent}
\end{figure}

This section investigates the cross-image transferability of various methods under different VLMs. We use both short targets (\eg ``Yes''), longer phrases (\eg``too late'', ``I'm sorry'') and short sentences (\eg ``This image is in black'', ``Absolutely! I can help you with that''). As shown in Tab.~\ref{tab:different_target}, \ourmethod consistently outperforms competing methods \textit{across all evaluated datasets, models, and attack targets}. Notably, it delivers an average relative improvement of over $30\%$ in ASR compared to SoTA methods. 
As shown in Fig.~\ref{time spent}, \ourmethod also reduces training times by approximately $70\%$ over white-box baselines while maintaining similar convergence, owing to its efficient optimization strategy that produces high-quality suffixes more quickly. This significant improvement demonstrates the superiority of \ourmethod, which performs end-to-end adversarial optimization of textual inputs in continuous space, guided by visual features. This allows \ourmethod to learn image-invariant features that enable the prompt to override visual grounding across images. 

\subsection{Black-box Transferability}
\label{cross-model}
We further evaluate our attack in a more challenging black-box scenario, where \ourmethod is transferred to attacker-unknown models. For the white-box attack methods, we employed a model ensemble approach to aggregate the gradients from MiniGPT-4 and BLIP-2 to generate adversarial prompts. For black-box methods \ct{TAP} and \ct{PAIR}, we use MiniGPT-4 as the target model. Fig.~\ref{Cross-Model Transferability}(c) indicates that our \ourmethod comprehensively outperforms all white-box baselines across the board. More critically, it proves superior on the majority of benchmarks ($2$ out of $3$) even when pitted against elite black-box methods (\textit{e.g.,} \ct{TAP} and \ct{PAIR}), which are designed for transferability. The results reveal that \ourmethod is also capable of generating transferable adversarial prompts across different VLMs, despite not being trained on the target models.

\begin{table}[t]
\centering
\small
\setlength{\tabcolsep}{1.5pt}
\renewcommand{\arraystretch}{0.85}
\caption{Effectiveness of our method and competitors against three prompt-level defenses. Our method is minimally affected by IP/RT and can bypass PF via AIR.}
\resizebox{\linewidth}{!}{%
\begin{tabular}{lccccc}
\toprule
\multirow{2}{*}{\textbf{Type}} & \multirow{2}{*}{\textbf{Method}}& \textbf{IP} &\textbf{RT} & \multicolumn{2}{c}{\textbf{PF}} \\
\cmidrule{5-6}
& & ASR & ASR & ASR & PPL \\
\midrule
\multirow{3}{*}{Black-box attack} 
& \ct{TAP} & 0.429 & 0.398 & 0.433 & 26.77 \\
& \ct{PAIR} & 0.395 & 0.400 & 0.402 & 31.90 \\
& \ct{BAP} & 0.311 & 0.305 & 0.299 & \textbf{25.15} \\
\midrule
\multirow{7}{*}{White-box attack} 
& \ct{AutoDAN} & 0.298 & 0.315 & 0.314 & 41.31 \\
& \ct{DeGCG} & 0.342 & 0.356 & 0.102 & 3012.35 \\
& \ct{Multi-I} & 0.280 & 0.296 & 0.071 & 3130.64 \\
& \ct{GCG-Transfer} & 0.274 & 0.295 & 0.059 & 3243.19 \\
& \ct{PGD-BERT} & 0.281 & 0.258 & 0.092 & 3115.32 \\
& \cellcolor{gray!8} \ourmethod &  \cellcolor{gray!8} \textbf{0.582} &  \cellcolor{gray!8} \textbf{0.542} &  \cellcolor{gray!8} 0.127 &  \cellcolor{gray!8} 864.98 \\
&  \cellcolor{gray!8} \ourmethod\hspace*{-0.3em}~\scriptsize{(+AIR)} &  \cellcolor{gray!8} 0.561 & \cellcolor{gray!8}  0.539 &  \cellcolor{gray!8} \textbf{0.459} &  \cellcolor{gray!8} 48.62 \\
\bottomrule
\end{tabular}
}
\label{defenses}
\end{table}

\section{Resistance to Potential Defenses}
\label{sec:defense}
Following~\citet{zhan2025adaptive,chang2025chatinject}, we evaluate \ourmethod against five representative defenses in MiniGPT-4: 
(1) \textit{Instructional Prevention}~\citep{learnprompting2024instructional_prevention}, 
(2) \textit{Retokenization}~\citep{jain2024baseline}, 
(3) \textit{Perplexity Filter}~\citep{alon2023detecting}, 
(4) \textit{Prompt Injection Detector}~\citep{deberta-v3-base-prompt-injection-v2}, and 
(5) \textit{Adversarial Training}~\citep{goodfellow2015explaining}. 
These defenses span a broad range of representative defense paradigms, enabling a stringent evaluation of our attack strength, \textit{including prompt-level defenses, perplexity-based filtering, detector-based methods~\citep{zeng2025psfd}, and training-based countermeasures}~\citep{yao2024reverse,zhang2025test}.

\subsection{Prompt-level Defenses}
We first evaluate two lightweight prompt-level defenses: \textit{Instructional Prevention} (IP)~\citep{learnprompting2024instructional_prevention}, which prepends warning instructions to encourage the model to ignore malicious content, and \textit{Retokenization} (RT)~\citep{jain2024baseline}, which perturbs the tokenization pattern of the input to disrupt suffixes that rely on specific token boundaries. Tab.~\ref{defenses} shows that both defenses reduce ASR by less than $10\%$, confirming their limited effectiveness against \ourmethod. This suggests that simple prompting or tokenization-level perturbation is insufficient to neutralize a transferable adversarial suffix optimized jointly with the target model.

\begin{figure}[t]
    \centering
    \subfigure[{Detector-based Defense}]{\includegraphics[width=0.32\linewidth]{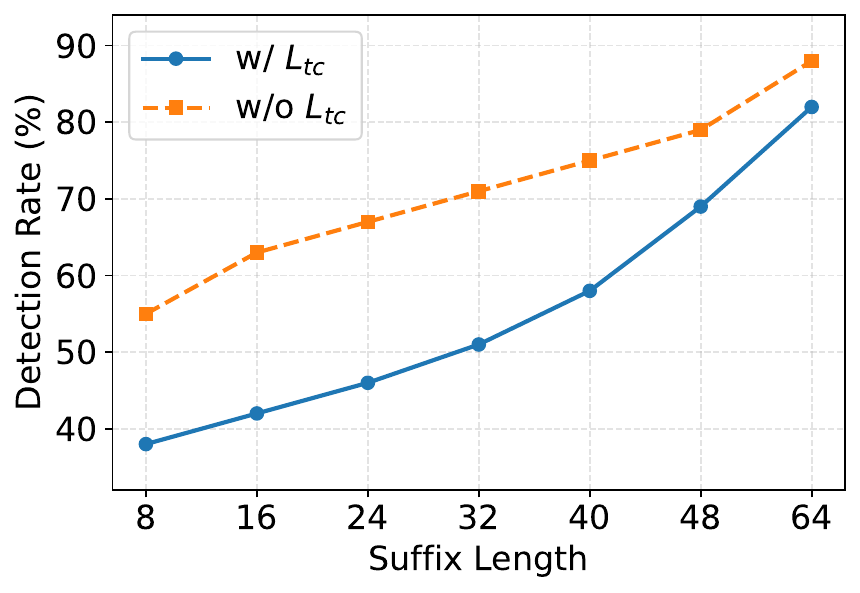}}
    \hfill
    \subfigure[{ Adversarial Training}]{\includegraphics[width=0.32\linewidth]{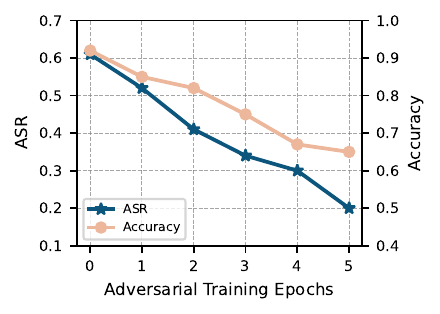}}
    \hfill
    \subfigure[{ Cross-Model}]{\includegraphics[width=0.33\linewidth]{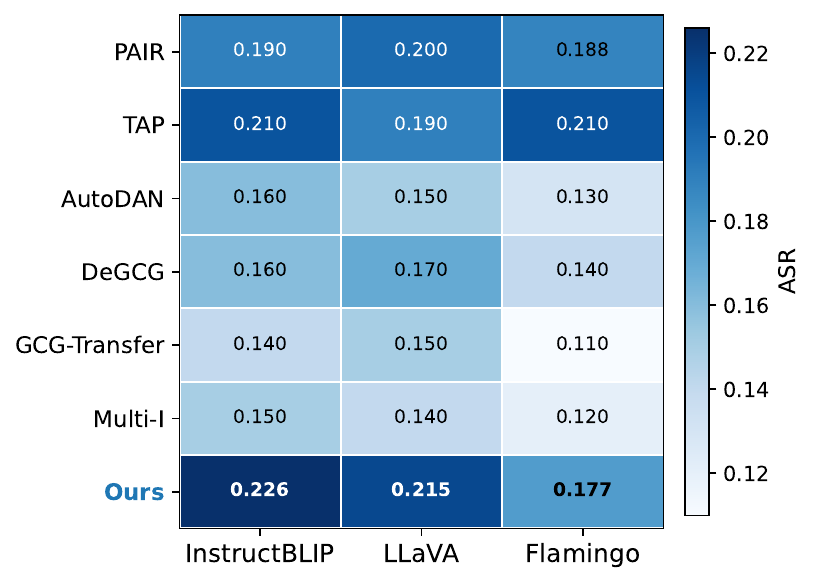}}
    \caption{(a) Detection rate under different suffix lengths, with and without $L_{tc}$. (b) ASR and Accuracy of MiniGPT-4 across adversarial training method. (c) Cross-model transferability of adversarial prompts.}
    \label{Cross-Model Transferability}
\end{figure}

\subsection{Perplexity-based Filtering}
We adopt the \textit{Perplexity Filter} (PF)~\citep{alon2023detecting}. 
As shown in Tab.~\ref{defenses}, almost all prompts with attack suffixes are flagged as abnormal by the filter. However, by leveraging the \textit{Adversarial Input Repetition} (AIR)~\citep{liao2024amplegcg} trick, repeating the prompt template 4 times before attaching the adversarial suffix, we observe that PPL is effectively reduced and leads to a recovery in ASR. Moreover, the PF leads to an unacceptably high false positive rate, significantly degrading user experience~\citep{liu2024autodan}. 

\subsection{Detector-based Defense}
We further evaluate a detector-based defense using the open-source \textit{Prompt Injection Detector}~\citep{deberta-v3-base-prompt-injection-v2}. We report the \ct{Detection Rate}, defined as the fraction of adversarial prompts that are correctly flagged as unsafe. As shown in Fig.~\ref{Cross-Model Transferability}(a), the detector can identify a portion of the attack prompts, but its effectiveness decreases when the suffix becomes shorter and more fluent. In particular, adding the Text Coherence Loss $\mathcal{L}_{\mathrm{tc}}$ reduces the detection rate by making the suffix more consistent with the model's next-token distribution, and reducing suffix length further weakens the surface evidence available to the detector. These results indicate that shorter and more fluent suffixes are harder for the detector to identify. 

\subsection{Adversarial Training}
Following~\citet{shao2024refusing}, we fine-tune MiniGPT-4 via LoRA~\citep{hu2022lowrank} using our attack samples. Specifically, we construct 100 prompt-image pairs by combining clean images with adversarial prompts generated by \ourmethod, while retaining the corresponding clean labels as targets.
As shown in Fig.~\ref{Cross-Model Transferability}(b), adversarial training consistently lowers the ASR of \ourmethod, reducing it from about $0.60$ without adversarial training to about $0.20$ after five training epochs. However, this robustness gain is accompanied by a substantial drop in clean-task accuracy~\citep{wang2026dual,wang2024unlearnable,li2025fine} (we sample visual-question pairs on VQAv2 dataset~\citep{goyal2017making}), which decreases from about $0.92$ to about $0.65$ over the same range of epochs. The trend therefore, reveals a clear trade-off in our setting: stronger robustness against adversarial suffixes comes at the cost of degraded performance on benign inputs. 

\begin{figure}[t]
    \centering
    \subfigure[{Efficacy of ``worst-case image''}]{\includegraphics[width=0.49\linewidth]{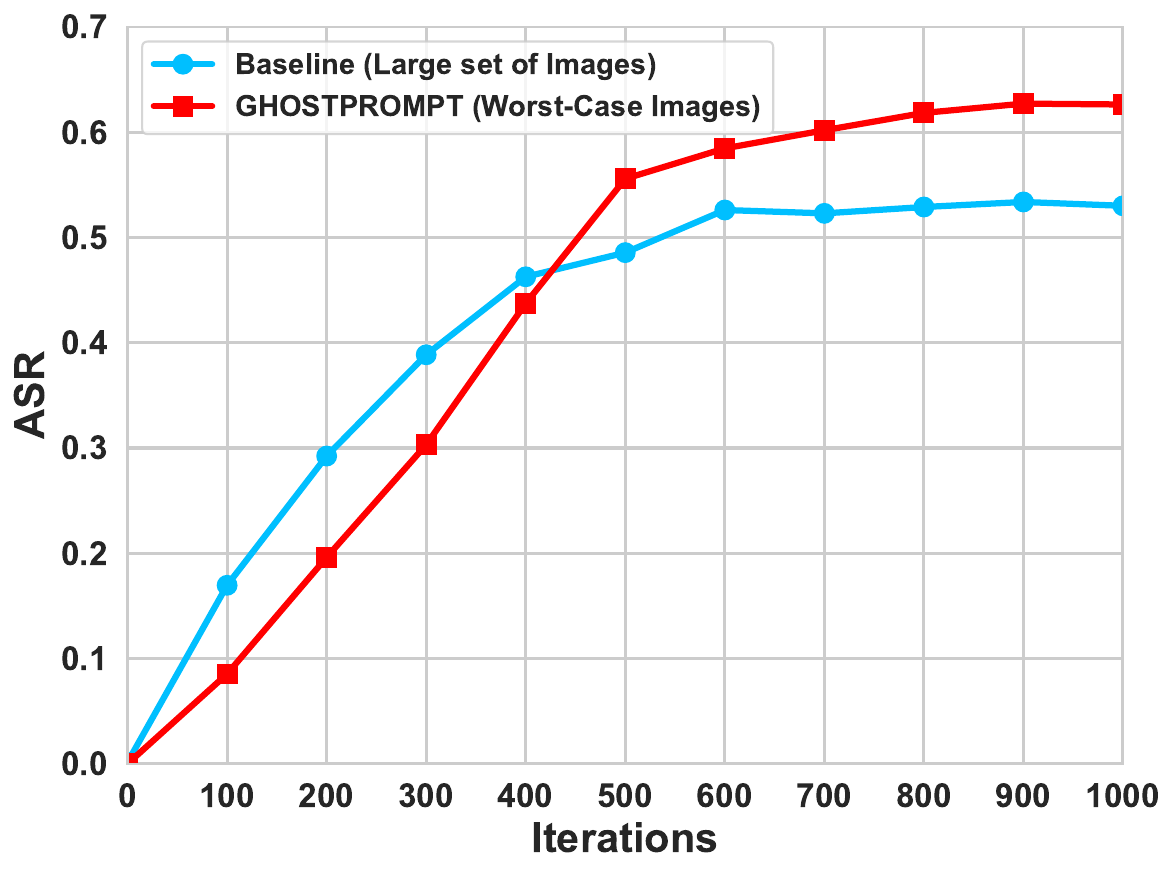}}
    \hfill
    \subfigure[{Attention mechanism}]{\includegraphics[width=0.49\linewidth]{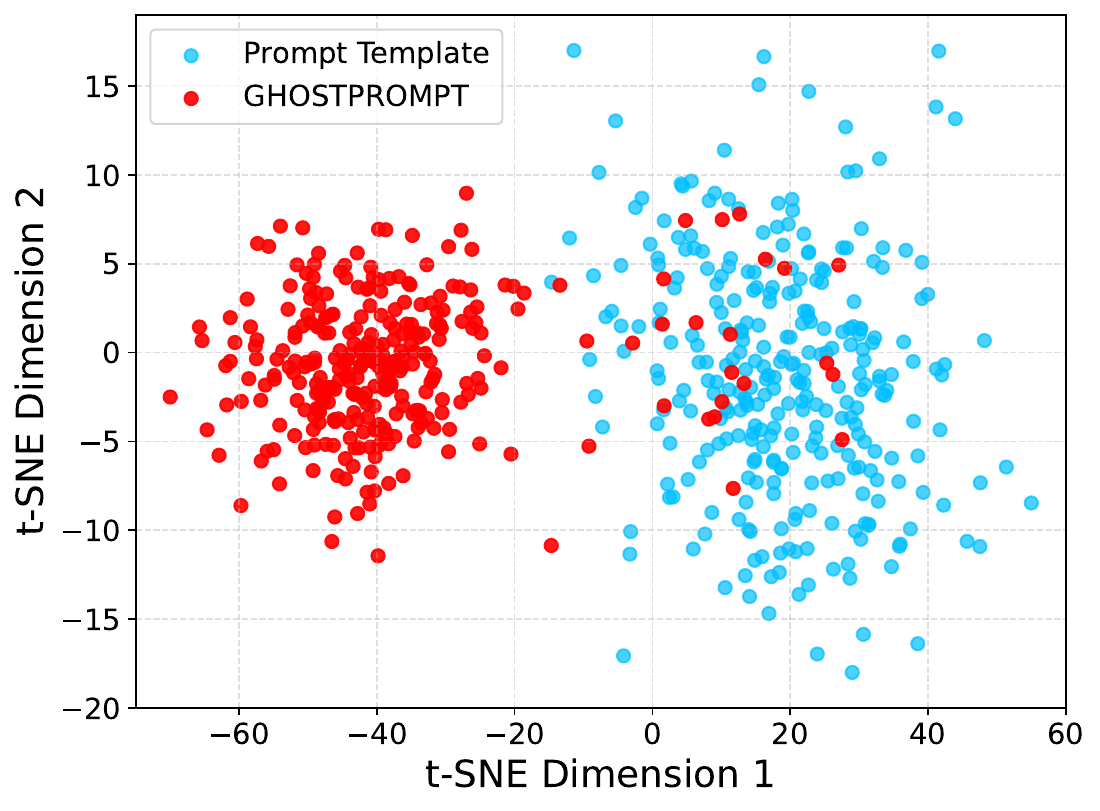}}
    \caption{Experimental validation of \ourmethod. (a) ASR for \ourmethod versus a random-image baseline over optimization iterations. (b) t-SNE visualization of internal response states under GhostPrompt and the prompt template.}
    \label{explain}
\end{figure}

\begin{table}[t]
\centering
\small
\setlength{\tabcolsep}{1.5pt}
\renewcommand{\arraystretch}{0.82}\
\caption{Effect of two key loss functions on \ourmethod on three VLMs, measured by ASR. Best performance is in bold.}
\resizebox{\linewidth}{!}{
\begin{tabular}{lccc}
\toprule
\textbf{Method}& \multicolumn{1}{c}{\textbf{MiniGPT-4}}  & \multicolumn{1}{c}{\textbf{BLIP-2}} & \multicolumn{1}{c}{\textbf{InstructBLIP}} \\
\midrule
\ct{w/o $\mathcal{L}_{\text{tc}}$} & 0.616 & \textbf{0.613} & \textbf{0.521} \\
\ct{w/o $\mathcal{L}_{\text{align}}$} & 0.558 & 0.541  & 0.447 \\
\ct{w/o $\mathcal{L}_{\text{tc}}$\;and\;$\mathcal{L}_{\text{align}}$} & 0.561 & 0.551 & 0.484  \\
 \cellcolor{gray!8} \ourmethod &  \cellcolor{gray!8}\textbf{0.620} &  \cellcolor{gray!8}0.597 &  \cellcolor{gray!8}0.512 \\
\bottomrule
\end{tabular}
}
\label{Ablation Study}
\end{table}

\section{A Closer Look at the Effectiveness of \ourmethod}
\label{effective}
We further analyze why \ourmethod is effective in the cross-image setting. Particularly, we examine two aspects of the method: (1) whether the proposed ``worst-case image'' optimization contributes to cross-image transferability, and (2) how the suffix affects the model's response patterns across diverse image inputs.

\noindent\textbf{``Worst-Case Image'' Optimization.}
We evaluate the contribution of the proposed ``worst-case image'' strategy by comparing \ourmethod with a strong data-augmentation baseline that optimizes the adversarial suffix using a large set of training images (1,000 images). As shown in Fig.~\ref{explain}(a), the data-augmentation baseline improves quickly at the beginning, but soon plateaus at a relatively low ASR, suggesting that simply exposing the optimization to more images is still insufficient to produce strong cross-image transferability. In contrast, \ourmethod exhibits a slower but more stable optimization trajectory and eventually converges to a substantially higher ASR. This result indicates that the gain of \ourmethod does not come merely from using more training images, but from explicitly constructing hard visual conditions during optimization.

\noindent\textbf{Analysis of Cross-Modal Response Patterns.}
To reveal the mechanism behind our attack, we use t-SNE~\citep{van2008visualizing} to visualize the model's internal states (represented by top-5 activation values of the first generated token) when processing diverse images. The results in Fig.~\ref{explain}(b) are striking: while the prompt template yields a scattered distribution (blue) that adapts to image content, \ourmethod forces these states into a tight, consistent cluster (red). This visually confirms the effectiveness of \ourmethod's \textbf{cross-image transferability}: it forges a universal, image-agnostic state by hijacking the model's decision-making process, compelling it to prioritize the text over any visual input.

\section{Ablation Study}
We systematically dissect the individual contributions of \ourmethod's components through controlled ablation experiments on MiniGPT-4 using the target response ``\textit{Yes}'' under the subset of the MS-COCO. 

\noindent\textbf{Impact of Different Loss.} 
We evaluate each loss component, with results in Tab.~\ref{Ablation 
Study}. Removing the text coherence loss leads to a slight increase in ASR, but at the cost of reduced fluency and naturalness of the generated suffix. In contrast, incorporating the semantic alignment loss improves ASR by an average of $7\%$, showing that it plays a more direct role in enhancing attack effectiveness.

\begin{figure}[t]
    \centering
    \subfigure[{ MiniGPT-4}]{\includegraphics[width=0.32\linewidth]{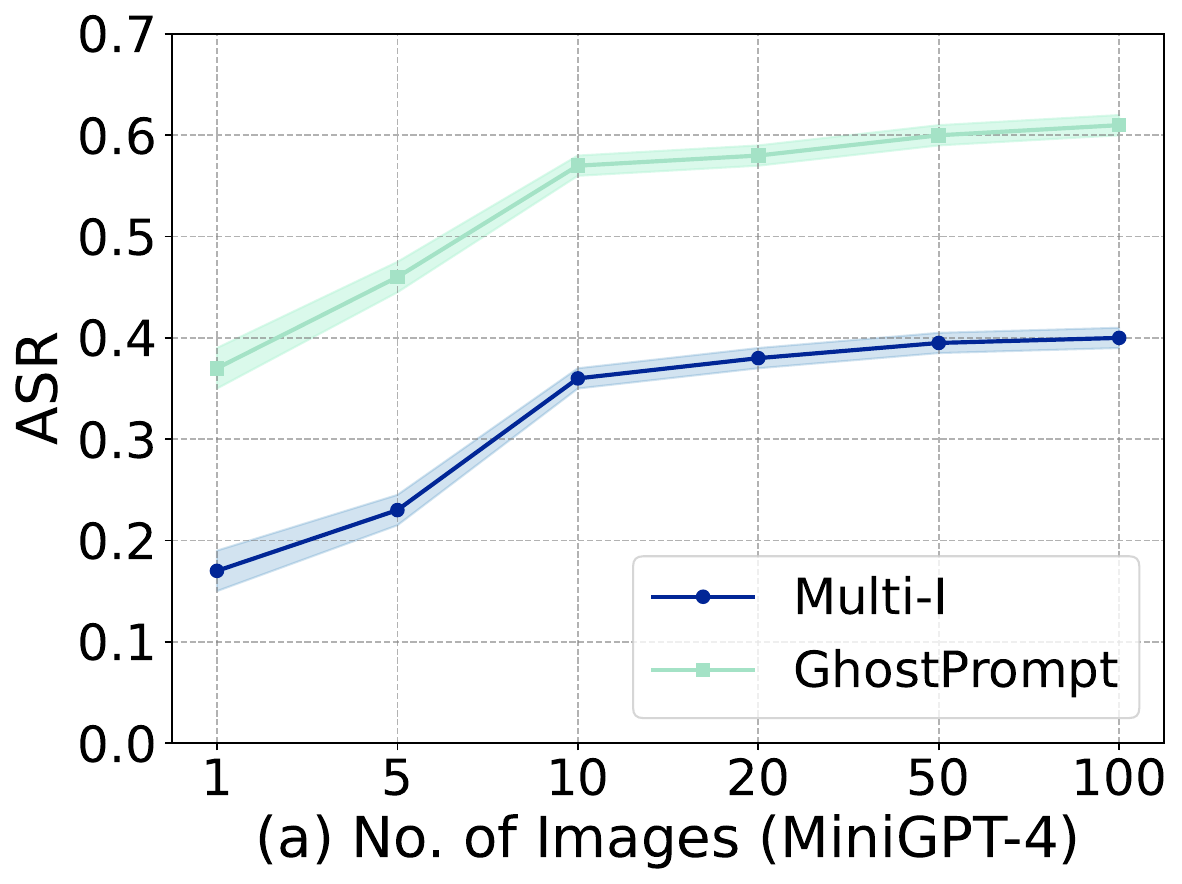}}
    \hfill
    \subfigure[{ BLIP-2}]{\includegraphics[width=0.32\linewidth]{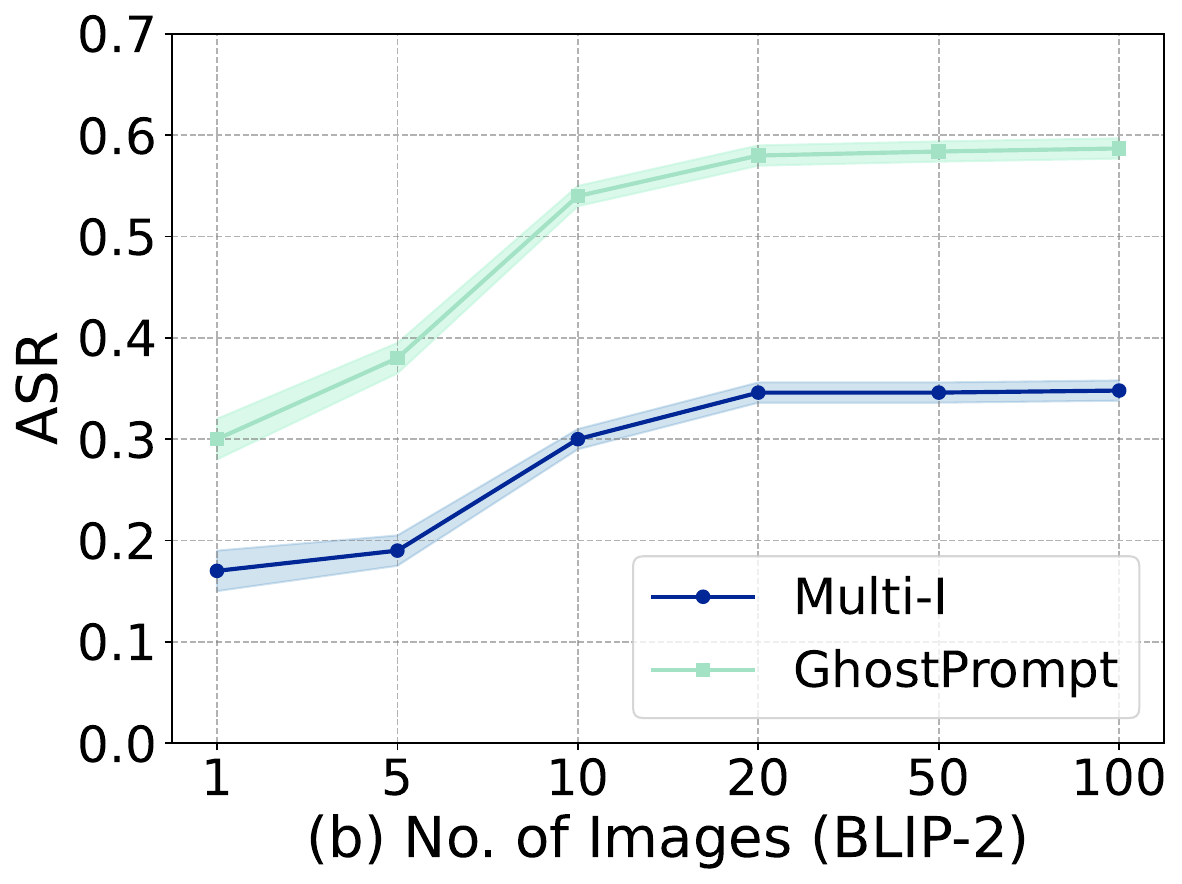}}
    \hfill
    \subfigure[{ InstructBLIP}]{\includegraphics[width=0.32\linewidth]{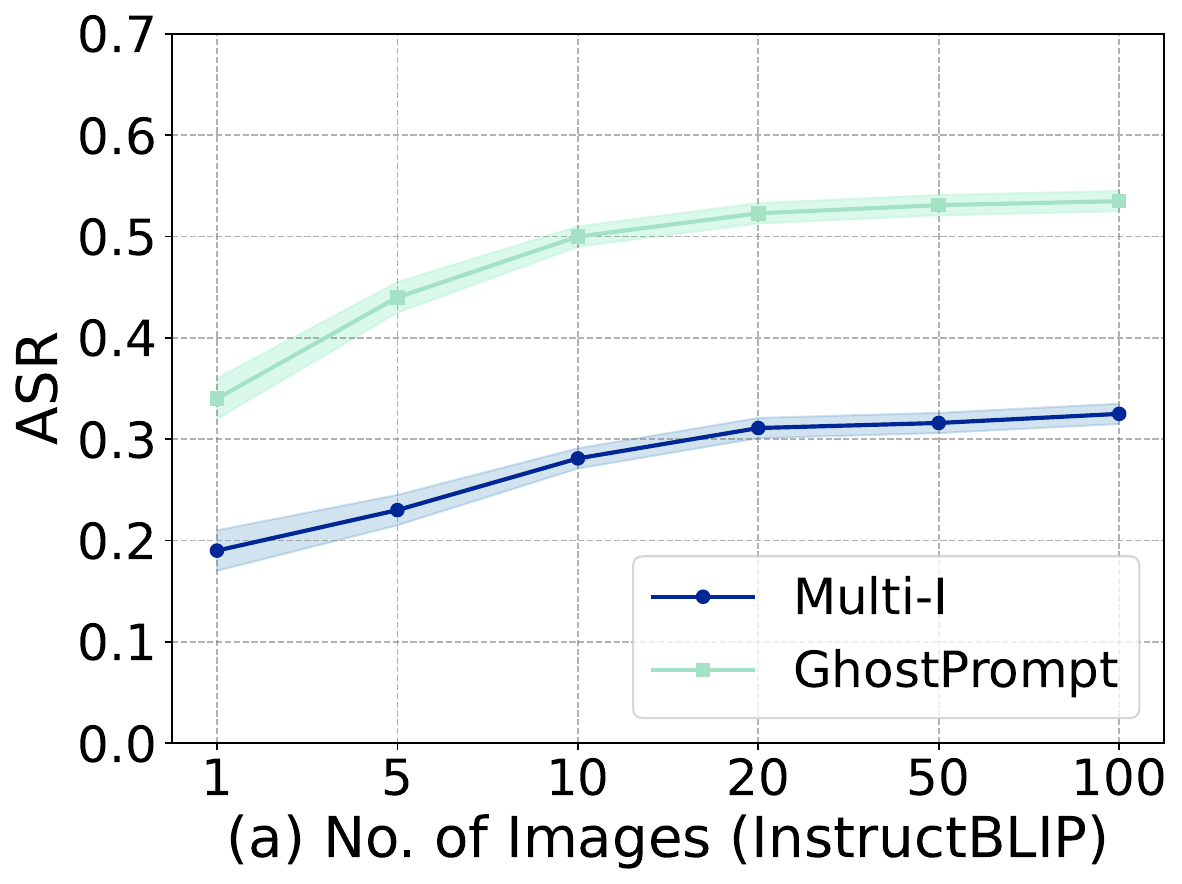}}
    \caption{ASR comparison between \ourmethod and \ct{Multi-I} when varying the numbers of the training image. We evaluate both methods on three VLMs using $1$, $5$, $10$, $20$, $50$ and $100$ training images.}
    \vspace{-1mm}
    \label{nums of image}
\end{figure}

\noindent\textbf{Number of images.} 
We study the effect of the number of training images used during optimization. As shown in Fig.~\ref{nums of image}, the ASR of both \ourmethod and \ct{Multi-I} increases at first and then becomes stable when enough training images are used. Based on this observation, we set the default number of training images to $20$. Moreover, \ourmethod consistently outperforms \ct{Multi-I} by more than $10\%$, demonstrating the advantage of our optimization strategy over straightforward multi-image training.


\section{Conclusion, Limitations, and Future Work}
In this paper, we present \ourmethod, a method for learning adversarial prompts with improved cross-image transferability. These findings highlight an underexplored robustness issue in VLMs and suggest that text-based attacks can exhibit cross-image transferability. Its limitations are partly inherited from text-based attacks. First, the discrete nature of language poses a challenge in achieving a near-perfect ASR compared to continuous image perturbations. Second, the black-box transferability of optimized prompts across significantly different VLM architectures remains limited. Finally, performance on commercial VLMs (\textit{e.g.,} GPT-4V) remains limited, which we leave for future work.

\section{Acknowledgment}
This work was supported in part by the Natural Science Foundation of China under Grant 62532009, 62372395, the Provincial Natural Science Foundation of Hunan under Grant No. 2025JJ50349, and the National Natural Science Foundation of China under Grant No. W2411053.

{
\bibliographystyle{ACM-Reference-Format}
\bibliography{ref}

@inproceedings{zhu2024minigpt,
  title = {Mini{GPT}-4: Enhancing Vision-Language Understanding with Advanced Large Language Models},
  author = {Deyao Zhu and Jun Chen and Xiaoqian Shen and Xiang Li and Mohamed Elhoseiny},
  booktitle = {Proceedings of the 12th International Conference on Learning Representations (ICLR'24)},
  pages = {1--17},
  year = {2024},
}

@article{pan2026towards,
  title={Towards Long-Horizon Interpretability: Efficient and Faithful Multi-Token Attribution for Reasoning LLMs},
  author={Pan, Wenbo and Liu, Zhichao and Wang, Xianlong and Yu, Haining and Jia, Xiaohua},
  journal={arXiv preprint arXiv:2602.01914},
  year={2026}
}

@article{zhou2024darksam,
  title={Darksam: Fooling segment anything model to segment nothing},
  author={Zhou, Ziqi and Song, Yufei and Li, Minghui and Hu, Shengshan and Wang, Xianlong and Zhang, Leo Yu and Yao, Dezhong and Jin, Hai},
  journal={Advances in Neural Information Processing Systems},
  volume={37},
  pages={49859--49880},
  year={2024}
}

@article{li2024safety,
  title={Safety layers in aligned large language models: The key to llm security},
  author={Li, Shen and Yao, Liuyi and Zhang, Lan and Li, Yaliang},
  journal={arXiv preprint arXiv:2408.17003},
  year={2024}
}

@article{gao2024shaping,
  title={Shaping the safety boundaries: Understanding and defending against jailbreaks in large language models},
  author={Gao, Lang and Geng, Jiahui and Zhang, Xiangliang and Nakov, Preslav and Chen, Xiuying},
  journal={arXiv preprint arXiv:2412.17034},
  year={2024}
}

@inproceedings{li2023blip2,
  title = {{BLIP}-2: Bootstrapping Language-Image Pre-training with Frozen Image Encoders and Large Language Models},
  author = {Junnan Li and Dongxu Li and Silvio Savarese and Steven Hoi},
  booktitle = {Proceedings of the 40th International Conference on Machine Learning (ICML'23)},
  pages = {12888--12900},
  year = {2023}
}

@inproceedings{goodfellow2015explaining,
  title = {Explaining and Harnessing Adversarial Examples},
  author = {Ian J. Goodfellow and Jonathon Shlens and Christian Szegedy},
  booktitle = {Proceedings of the 3rd International Conference on Learning Representations (ICLR'15)},
  pages = {1--11},
  year = {2015}
}

@inproceedings{luo2024image,
  title = {An Image Is Worth 1000 Lies: Transferability of Adversarial Images across Prompts on Vision-Language Models},
  author = {Luo, Haochen and Gu, Jindong and Liu, Fengyuan and Torr, Philip},
  booktitle = {Proceedings of the 12th International Conference on Learning Representations (ICLR'24)},
  pages = {1--15},
  year = {2024},
}

@article{zou2023universal,
  title={Universal and transferable adversarial attacks on aligned language models},
  author={Zou, Andy and Wang, Zifan and Carlini, Nicholas and Nasr, Milad and Kolter, J Zico and Fredrikson, Matt},
  journal={arXiv preprint arXiv:2307.15043},
  year={2023}
}

@inproceedings{zhang2025anyattack,
  title={Anyattack: Towards Large-scale Self-supervised Adversarial Attacks on Vision-language Models},
  author={Zhang, Jiaming and Ye, Junhong and Ma, Xingjun and Li, Yige and Yang, Yunfan and Chen, Yunhao and Sang, Jitao and Yeung, Dit-Yan},
  booktitle={Proceedings of the 2025 Computer Vision and Pattern Recognition Conference (CVPR'25)},
  pages={19900--19909},
  year={2025}
}

@inproceedings{wang2024white,
  title = {White-box Multimodal Jailbreaks Against Large Vision-Language Models},
  author = {Wang, Ruofan and Ma, Xingjun and Zhou, Hanxu and Ji, Chuanjun and Ye, Guangnan and Jiang, Yu-Gang},
  booktitle = {Proceedings of the 32nd ACM International Conference on Multimedia (MM'24)},
  pages = {6920--6928},
  year = {2024},
  publisher = {Association for Computing Machinery},
}

@inproceedings{szegedy2014intriguing,
  title={Intriguing properties of neural networks},
  author={Christian Szegedy and Wojciech Zaremba and Ilya Sutskever and Joan Bruna and Dumitru Erhan and Ian Goodfellow and Rob Fergus},
  booktitle={Proceedings of the 2nd International Conference on Learning Representations (ICLR'14)},
  year={2014}
}

@inproceedings{moosavi2017universal,
  title={Universal adversarial perturbations},
  author={Seyed-Mohsen Moosavi-Dezfooli and Alhussein Fawzi and Omar Fawzi and Pascal Frossard},
  booktitle={Proceedings of the 2017 IEEE/CVF Conference on Computer Vision and Pattern Recognition (CVPR'17)},
  pages={86--94},
  year={2017}
}

@inproceedings{madry2018towards,
  title={Towards deep learning models resistant to adversarial attacks},
  author={Aleksander Madry and Aleksandar Makelov and Ludwig Schmidt and Dimitris Tsipras and Adrian Vladu},
  booktitle={Proceedings of the 6th International Conference on Learning Representations (ICLR'18)},
  year={2018}
}

@inproceedings{dai2023instructblip,
  title={InstructBLIP: Towards General-purpose Vision-Language Models with Instruction Tuning},
  author={Wenliang Dai and Junnan Li and Dongxu Li and Anthony Meng Huat Tiong and Junqi Zhao and Weisheng Wang and Boyang Albert Li and Pascale Fung and Steven C. H. Hoi},
  booktitle={Proceedings of the 37th International Conference on Neural Information Processing Systems (NeurIPS'23)},
  year={2023},
  volume={36}
}

@inproceedings{jang2017categorical,
  title={Categorical reparameterization with gumbel-softmax},
  author={Eric Jang and Shixiang Gu and Ben Poole},
  booktitle={Proceedings of the 5th International Conference on Learning Representations (ICLR'17)},
  year={2017}
}

@inproceedings{qi2024visual,
  title={Visual Adversarial Examples Jailbreak Aligned Large Language Models},
  author={Qi, Xiangyu and Huang, Kaixuan and Panda, Ashwinee and Henderson, Peter and Wang, Mengdi and Mittal, Prateek},
  booktitle={Proceedings of the 38th AAAI Conference on Artificial Intelligence (AAAI'24)},
  volume={38},
  number={19},
  pages={21527--21536},
  year={2024},
}

@inproceedings{lin2014microsoft,
  title = {Microsoft COCO: Common Objects in Context},
  author = {Lin, Tsung-Yi and Maire, Michael and Belongie, Serge and Hays, James and Perona, Pietro and Ramanan, Deva and Doll\'ar, Piotr and Zitnick, C. Lawrence},
  booktitle = {Proceedings of the 13th European Conference on Computer Vision (ECCV'14)},
  pages = {740--755},
  year = {2014}
}

@inproceedings{ouyang2022training,
  title = {Training Language Models to Follow Instructions with Human Feedback},
  author    = {Ouyang, Long and Wu, Jeffrey and Jiang, Xu and Almeida, Diogo and Wainwright, Carroll and Mishkin, Pamela and Zhang, Chong and Agarwal, Sandhini and Slama, Katarina and others},
  booktitle = {Proceedings of the 36th International Conference on Neural Information Processing Systems (NeurIPS'22)},
  volume = {35},
  pages = {27730--27744},
  year = {2022}
}

@article{liao2024amplegcg,
  title={Amplegcg: Learning a universal and transferable generative model of adversarial suffixes for jailbreaking both open and closed llms},
  author={Liao, Zeyi and Sun, Huan},
  journal={arXiv preprint arXiv:2404.07921},
  year={2024}
}

@inproceedings{radford2021learning,
  title={Learning Transferable Visual Models From Natural Language Supervision},
  author={Radford, Alec and Kim, Jong Wook and Hallacy, Chris and Ramesh, Aditya and Goh, Gabriel and Agarwal, Sandhini and Sastry, Girish and Askell, Amanda and Mishkin, Pamela and Clark, Jack and Krueger, Gretchen and Sutskever, Ilya},
  booktitle={Proceedings of the 38th International Conference on Machine Learning (ICML'21)},
  pages={8748--8763},
  year={2021},
  volume={139},
  publisher={PMLR}
}

@inproceedings{liu2024autodan,
  title={Auto{DAN}: Generating Stealthy Jailbreak Prompts on Aligned Large Language Models},
  author={Liu, Xiaogeng and Xu, Nan and Chen, Muhao and Xiao, Chaowei},
  booktitle={Proceedings of the 12th International Conference on Learning Representations (ICLR'24)},
  year={2024},
}

@article{chao2023jailbreaking,
  title={Jailbreaking Black Box Large Language Models in Twenty Queries},
  author={Chao, Patrick and Robey, Alexander and Dobriban, Edgar and Hassani, Hamed and Pappas, George J and Wong, Eric},
  journal={arXiv preprint arXiv:2310.08419},
  year={2023}
}

@misc{promptbase,
  title = {{PromptBase} | Prompt Marketplace: {Midjourney}, {ChatGPT}, {Sora}, {FLUX} \& more},
  author = {{PromptBase}},
  url = {https://promptbase.com/},
  year = {2025},
  note = {Accessed: 2025-07-15}
}

@article{mehrotra2023treeOfAttacks,
  title={Tree of Attacks: Jailbreaking Black-Box LLMs Automatically},
  author={Mehrotra, Anay and Zampetakis, Manolis and Kassianik, Paul and Nelson, Blaine and Anderson, Hyrum and Singer, Yaron and Karbasi, Amin},
  journal={arXiv preprint arXiv:2312.02119},
  year={2023}
}

@article{ying2024jailbreak,
  title={Jailbreak Vision Language Models via Bi-Modal Adversarial Prompt},
  author={Ying, Zonghao and Liu, Aishan and Zhang, Tianyuan and Yu, Zhengmin and Liang, Siyuan and Liu, Xianglong and Tao, Dacheng},
  journal={arXiv preprint arXiv:2406.04031},
  year={2024}
}

@inproceedings{jain2024baseline,
  title = {Baseline Defenses for Adversarial Attacks Against Aligned Language Models},
  author = {Jain, Neel and Schwarzschild, Avi and Wen, Yuxin and Somepalli, Gowthami and Kirchenbauer, John and Chiang, Ping‑yeh and Goldblum, Micah and Saha, Aniruddha and Geiping, Jonas and Goldstein, Tom},
  booktitle = {Proceedings of the 12th International Conference on Learning Representations (ICLR'24)},
  year = {2024}
}

@article{alon2023detecting,
  title={Detecting language model attacks with perplexity},
  author={Alon, Gabriel and Kamfonas, Michael},
  journal={arXiv preprint arXiv:2308.14132},
  year={2023}
}

@inproceedings{liu2024advancing,
  title = {Advancing Adversarial Suffix Transfer Learning on Aligned Large Language Models},
  author = {Liu, Hongfu and Xie, Yuxi and Wang, Ye and Shieh, Michael},
  booktitle = {Proceedings of the 29th Conference on Empirical Methods in Natural Language Processing (EMNLP'24)},
  year = {2024},
  pages = {7213--7224},
}

@article{russakovsky2015imagenet,
  author  = {Russakovsky, Olga and Deng, Jia and Su, Hao and Krause, Jonathan and
             Satheesh, Sanjeev and Ma, Sean and Huang, Zhiheng and Karpathy, Andrej and
             Khosla, Aditya and Bernstein, Michael S. and Berg, Alexander C. and
             Fei-Fei, Li},
  title   = {ImageNet Large Scale Visual Recognition Challenge},
  journal = {International Journal of Computer Vision},
  volume  = {115},
  number  = {3},
  pages   = {211--252},
  year    = {2015},
}

@article{touvron2023llama2,
  title={Llama 2: Open foundation and fine-tuned chat models},
  author={Touvron, Hugo and Martin, Louis and Stone, Kevin and Albert, Peter and Almahairi, Amjad and Babaei, Yasmine and Bashlykov, Nikolay and Batra, Soumya and Bhargava, Prajjwal and Bhosale, Shruti and others},
  journal={arXiv preprint arXiv:2307.09288},
  year={2023}
}

@article{achiam2023gpt4,
  title={Gpt-4 technical report},
  author={Achiam, Josh and Adler, Steven and Agarwal, Sandhini and Ahmad, Lama and Akkaya, Ilge and Aleman, Florencia Leoni and Almeida, Diogo and Altenschmidt, Janko and Altman, Sam and Anadkat, Shyamal and others},
  journal={arXiv preprint arXiv:2303.08774},
  year={2023}
}

@inproceedings{zhao2023onevaluating,
 author = {Zhao, Yunqing and Pang, Tianyu and Du, Chao and Yang, Xiao and LI, Chongxuan and Cheung, Ngai-Man (Man) and Lin, Min},
 booktitle = {Proceedings of the 37th Conference on Neural Information Processing Systems (NeurIPS'23)},
 pages = {54111--54138},
 title = {On Evaluating Adversarial Robustness of Large Vision-Language Models},
 volume = {36},
 year = {2023}
}

@article{brown2017adversarial,
  title={Adversarial patch},
  author={Brown, Tom B and Man{\'e}, Dandelion and Roy, Aurko and Abadi, Mart{\'\i}n and Gilmer, Justin},
  journal={arXiv preprint arXiv:1712.09665},
  year={2017}
}

@inproceedings{zhan2025adaptive,
    title = "Adaptive Attacks Break Defenses Against Indirect Prompt Injection Attacks on {LLM} Agents",
    author = "Zhan, Qiusi and Fang, Richard and Panchal, Henil Shalin and Kang, Daniel",
    booktitle = "Proceedings of the 2025 Conference of the North {A}merican Chapter of the Association for Computational Linguistics: Human Language Technologies (NAACL'25)",
    year = "2025",
    pages = "7101--7117",
}

@inproceedings{hu2022lowrank,
  title     = {LoRA: Low‑Rank Adaptation of Large Language Models},
  author    = {Hu, Edward J. and Shen, Yelong and Wallis, Phillip and Allen‑Zhu, Zeyuan and Li, Yuanzhi and Wang, Shean and Wang, Lu and Chen, Weizhu},
  booktitle = {Proceedings of the 12th International Conference on Learning Representations (ICLR'22)},
  year      = {2022},
}

@article{shao2024refusing,
  title={Refusing safe prompts for multi-modal large language models},
  author={Shao, Zedian and Liu, Hongbin and Hu, Yuepeng and Gong, Neil Zhenqiang},
  journal={arXiv preprint arXiv:2407.09050},
  year={2024}
}

@inproceedings{huang2025xtransfer,
  title     = {X‑Transfer Attacks: Towards Super Transferable Adversarial Attacks on {CLIP}},
  author    = {Hanxun Huang and Sarah Monazam Erfani and Yige Li and Xingjun Ma and James Bailey},
  booktitle = {Proceedings of the 42nd International Conference on Machine Learning (ICML'25)}, 
  year      = {2025},
  note      = {Poster presentation},
}

@inproceedings{lu2023setlevel,
  title     = {Set‑level Guidance Attack: Boosting Adversarial Transferability of Vision‑Language Pre‑training Models},
  author    = {Dong Lu and Zhiqiang Wang and Teng Wang and Weili Guan and Hongchang Gao and Feng Zheng},
  booktitle = {Proceedings of the IEEE/CVF International Conference on Computer Vision (ICCV’23)},
  pages     = {102--111},
  year      = {2023},
}

@inproceedings{goyal2017making,
  title     = {Making the V in VQA Matter: Elevating the Role of Image Understanding in Visual Question Answering},
  author    = {Goyal, Yash and Khot, Tejas and Summers‑Stay, Douglas and Batra, Dhruv and Parikh, Devi},
  booktitle = {Proceedings of the IEEE Conference on Computer Vision and Pattern Recognition (CVPR'17)},
  year      = {2017},
  pages     = {6904--6913},
}

@article{liu2023visual,
  title={Visual Instruction Tuning},
  author={Liu, Haotian and Li, Chunyuan and Wu, Qingyang and Lee, Yong Jae},
  journal={arXiv preprint arXiv:2304.08485},
  year={2023}
}

@article{van2008visualizing,
  title={Visualizing data using t-SNE},
  author={van der Maaten, Laurens and Hinton, Geoffrey},
  journal={Journal of Machine Learning Research},
  volume={9},
  pages={2579--2605},
  year={2008}
}

@article{waghela2024enhancing,
  title={Enhancing adversarial text attacks on bert models with projected gradient descent},
  author={Waghela, Hetvi and Sen, Jaydip and Rakshit, Sneha},
  journal={arXiv preprint arXiv:2407.21073},
  year={2024}
}

@article{yang2024enhancing,
  title={Enhancing Cross-Prompt Transferability in Vision-Language Models through Contextual Injection of Target Tokens},
  author={Yang, Xikang and Tang, Xuehai and Zhu, Fuqing and Han, Jizhong and Hu, Songlin},
  journal={arXiv preprint arXiv:2406.13294},
  year={2024}
}

@inproceedings{hudson2019gqa,
  title={GQA: A new dataset for real-world visual reasoning and compositional question answering},
  author={Hudson, Drew A and Manning, Christopher D},
  booktitle={Proceedings of the IEEE/CVF conference on computer vision and pattern recognition (CVPR'19)},
  pages={6700--6709},
  year={2019}
}

@misc{deberta-v3-base-prompt-injection-v2,
  author = {ProtectAI.com},
  title = {Fine-Tuned DeBERTa-v3-base for Prompt Injection Detection},
  year = {2024},
  publisher = {HuggingFace},
  url = {https://huggingface.co/ProtectAI/deberta-v3-base-prompt-injection-v2},
}

@misc{learnprompting2024instructional_prevention,
  author       = {Learn Prompting},
  title        = {Instruction Defense},
  howpublished = {\url{https://learnprompting.org/docs/prompt_hacking/defensive_measures/instruction}},
  year         = {2024}
}

@article{chang2025chatinject,
  title={Chatinject: Abusing chat templates for prompt injection in llm agents},
  author={Chang, Hwan and Jun, Yonghyun and Lee, Hwanhee},
  journal={arXiv preprint arXiv:2509.22830},
  year={2025}
}

@inproceedings{zhang2026defending,
  title     = {Defending Jailbreak Attacks on Large Language Models via Manifold Trajectory Kinetics},
  author    = {Zhang, Hangtao and Zhao, Yucheng and Liu, Sishun and Zhou, Ziqi and Ye, Zeyu and Wan, Wei and Li, Minghui and Hu, Shengshan and Zhang, Yanjun and Liu, Yi and Zhang, Leo Yu},
  booktitle = {35th USENIX Security Symposium (USENIX Security 26)},
  year      = {2026}
}

@inproceedings{zeng2025psfd,
  title={PSFD: Proactive Spatial-Frequency Defense against Malicious Exemplar-Guided Image Editing},
  author={Zeng, Li and Mo, Xiaojun and Xie, Meng and Zhang, Hangtao and Liu, Yixiang and Peng, Yezhuo and Li, Yanchun},
  booktitle={2025 IEEE International Conference on Multimedia and Expo (ICME)},
  pages={1--6},
  year={2025},
  organization={IEEE}
}

@article{wang2026dual,
  title={Dual-branch Robust Unlearnable Examples},
  author={Wang, Xianlong and Zhang, Hangtao and Pan, Wenbo and Zhou, Ziqi and Jiang, Changsong and Zeng, Li and Jia, Xiaohua},
  journal={arXiv preprint arXiv:2605.01718},
  year={2026}
}

@article{li2025fine,
  title={Fine-grained poisoning framework against federated learning},
  author={Li, Minghui and Zhang, Hangtao and Zhang, Yanjun and Zeng, Li and Chen, Chao and Shao, Qiyun and Wan, Wei and Hu, Shengshan and Zhang, Leo Yu},
  journal={IEEE Transactions on Dependable and Secure Computing},
  year={2025},
  publisher={IEEE}
}

@article{song2026segtrans,
  title={Segtrans: Transferable adversarial examples for segmentation models},
  author={Song, Yufei and Zhou, Ziqi and Lu, Qi and Zhang, Hangtao and Hu, Yifan and Xue, Lulu and Hu, Shengshan and Li, Minghui and Zhang, Leo Yu},
  journal={IEEE Transactions on Multimedia},
  year={2026},
  publisher={IEEE}
}

@article{zhou2025darkhash,
  title={Darkhash: A data-free backdoor attack against deep hashing},
  author={Zhou, Ziqi and Deng, Menghao and Song, Yufei and Zhang, Hangtao and Wan, Wei and Hu, Shengshan and Li, Minghui and Zhang, Leo Yu and Yao, Dezhong},
  journal={IEEE Transactions on Information Forensics and Security},
  year={2025},
  publisher={IEEE}
}

@article{wang2026advedm,
  title={Advedm: Fine-grained adversarial attack against vlm-based embodied agents},
  author={Wang, Yichen and Zhang, Hangtao and Pan, Hewen and Zhou, Ziqi and Wang, Xianlong and Guo, Peijin and Xue, Lulu and Hu, Shengshan and Li, Minghui and Zhang, Leo Yu},
  journal={Advances in Neural Information Processing Systems},
  volume={38},
  pages={136551--136575},
  year={2026}
}

@inproceedings{wang2025breakingphysical,
  title={Breaking barriers in physical-world adversarial examples: Improving robustness and transferability via robust feature},
  author={Wang, Yichen and Chou, Yuxuan and Zhou, Ziqi and Zhang, Hangtao and Wan, Wei and Hu, Shengshan and Li, Minghui},
  booktitle={Proceedings of the AAAI Conference on Artificial Intelligence},
  volume={39},
  number={8},
  pages={8069--8077},
  year={2025}
}

@article{wang2024unlearnable,
  title={Unlearnable 3d point clouds: Class-wise transformation is all you need},
  author={Wang, Xianlong and Li, Minghui and Liu, Wei and Zhang, Hangtao and Hu, Shengshan and Zhang, Yechao and Zhou, Ziqi and Jin, Hai},
  journal={Advances in Neural Information Processing Systems},
  volume={37},
  pages={99404--99432},
  year={2024}
}

@inproceedings{song2025pb,
  title={Pb-uap: Hybride universal adversarial attack for image segmentation},
  author={Song, Yufei and Zhou, Ziqi and Li, Minghui and Wang, Xianlong and Zhang, Hangtao and Deng, Menghao and Wan, Wei and Hu, Shengshan and Zhang, Leo Yu},
  booktitle={ICASSP 2025-2025 IEEE International Conference on Acoustics, Speech and Signal Processing (ICASSP)},
  pages={1--5},
  year={2025},
  organization={IEEE}
}

@article{yao2024reverse,
  title={Reverse backdoor distillation: Towards online backdoor attack detection for deep neural network models},
  author={Yao, Zeming and Zhang, Hangtao and Guo, Yicheng and Tian, Xin and Peng, Wei and Zou, Yi and Zhang, Leo Yu and Chen, Chao},
  journal={IEEE Transactions on Dependable and Secure Computing},
  volume={21},
  number={6},
  pages={5098--5111},
  year={2024},
  publisher={IEEE}
}

@inproceedings{zhang2025test,
  title={Test-time backdoor detection for object detection models},
  author={Zhang, Hangtao and Wang, Yichen and Yan, Shihui and Zhu, Chenyu and Zhou, Ziqi and Hou, Linshan and Hu, Shengshan and Li, Minghui and Zhang, Yanjun and Zhang, Leo Yu},
  booktitle={Proceedings of the Computer Vision and Pattern Recognition Conference},
  pages={24377--24386},
  year={2025}
}

@article{zhang2023denial,
  title={Denial-of-service or fine-grained control: Towards flexible model poisoning attacks on federated learning},
  author={Zhang, Hangtao and Yao, Zeming and Zhang, Leo Yu and Hu, Shengshan and Chen, Chao and Liew, Alan and Li, Zhetao},
  journal={arXiv preprint arXiv:2304.10783},
  year={2023}
}

@article{zhang2024detector,
  title={Detector collapse: Physical-world backdooring object detection to catastrophic overload or blindness in autonomous driving},
  author={Zhang, Hangtao and Hu, Shengshan and Wang, Yichen and Zhang, Leo Yu and Zhou, Ziqi and Wang, Xianlong and Zhang, Yanjun and Chen, Chao},
  journal={arXiv preprint arXiv:2404.11357},
  year={2024}
}

@article{wang2024trojanrobot,
  title={Trojanrobot: Physical-world backdoor attacks against vlm-based robotic manipulation},
  author={Wang, Xianlong and Pan, Hewen and Zhang, Hangtao and Li, Minghui and Hu, Shengshan and Zhou, Ziqi and Xue, Lulu and Liu, Aishan and Jiang, Yunpeng and Zhang, Leo Yu and others},
  journal={arXiv preprint arXiv:2411.11683},
  year={2024}
}

@inproceedings{zhangbadrobot,
  title={BadRobot: Jailbreaking Embodied LLM Agents in the Physical World},
  author={Zhang, Hangtao and Zhu, Chenyu and Wang, Xianlong and Zhou, Ziqi and Yin, Changgan and Li, Minghui and Xue, Lulu and Wang, Yichen and Hu, Shengshan and Liu, Aishan and others},
  booktitle={Proceedings of the Thirteenth International Conference on Learning Representations},
  year={2025}
}

@inproceedings{zhou2023advclip,
  title={Advclip: Downstream-agnostic adversarial examples in multimodal contrastive learning},
  author={Zhou, Ziqi and Hu, Shengshan and Li, Minghui and Zhang, Hangtao and Zhang, Yechao and Jin, Hai},
  booktitle={Proceedings of the 31st ACM International Conference on Multimedia},
  pages={6311--6320},
  year={2023}
}

@article{li2026privacy,
  title={Privacy-Preserving Yet Vulnerable: Data Poisoning Attacks Against Differential Privacy Sparse Mobile Crowdsensing System},
  author={Li, Chengxin and Gu, Yujie and Qiao, Pengpeng and Pan, Shengli and Sakurai, Kouichi and Li, Zhetao},
  journal={IEEE Transactions on Mobile Computing},
  year={2026},
  publisher={IEEE}
}

@article{chen2026tex3d,
  title={Tex3D: Objects as attack surfaces via adversarial 3D textures for vision-language-action models},
  author={Chen, Jiawei and Huang, Simin and Du, Jiawei and Chen, Shuaihang and Tian, Yu and Wei, Mingjie and Yu, Chao and Yin, Zhaoxia},
  journal={arXiv preprint arXiv:2604.01618},
  year={2026}
}

@inproceedings{chen2026red,
  title={Red teaming large reasoning models},
  author={Chen, Jiawei and Yang, Yang and Yu, Chao and Tian, Yu and Cao, Zhi and Yang, Xue and Li, Linghao and Su, Hang and Yin, Zhaoxia},
  booktitle={Proceedings of the 64th Annual Meeting of the Association for Computational Linguistics (Volume 1: Long Papers)},
  pages={22559--22591},
  year={2026}
}

@article{long2025fault,
  title={Fault-Tolerant Aware Task Offloading Based on Reinforcement Learning in Mobile Edge Computing},
  author={Long, Saiqin and Rao, Chongxi and Liu, Haolin and Chen, Yunjie and Deng, Qingyong and Shang, Jing and Li, Zhetao},
  journal={IEEE Transactions on Mobile Computing},
  year={2025},
  publisher={IEEE}
}
}
\end{document}